\documentclass[prodmode]{acmsmall} 

\providecommand{\tabularnewline}{\\}

\usepackage[ruled]{algorithm2e}
\usepackage{array}
\usepackage{multirow}
\usepackage{amsmath}
\usepackage{amssymb}
\usepackage{graphicx}

\SetAlFnt{\small}
\SetAlCapFnt{\small}
\SetAlCapNameFnt{\small}
\SetAlCapHSkip{0pt}
\IncMargin{-\parindent}

\begin{document}

\markboth{J. Pawlick et al.}{Defensive Deception for Cybersecurity and Privacy}

\title{A Game-Theoretic Taxonomy and Survey of Defensive Deception for Cybersecurity and Privacy}
\author{Jeffrey Pawlick
\affil{NYU Tandon School of Engineering, and U.S. Army Research Laboratory}
Edward Colbert
\affil{Virginia Tech Intelligent Systems Lab, Hume Center for National Security and Technology, and U.S. Army Research Laboratory}
Quanyan Zhu
\affil{New York University Tandon School of Engineering}
}

\begin{abstract}
Cyberattacks on both databases and critical infrastructure have threatened public and private sectors. Ubiquitous tracking and wearable computing have infringed upon privacy. Advocates and engineers have recently proposed using defensive deception as a means to leverage the information asymmetry typically enjoyed by attackers as a tool for defenders. The term deception, however, has been employed broadly and with a variety of meanings. In this paper, we survey 24 articles from 2008--2018 that use game theory to model defensive deception for cybersecurity and privacy. Then we propose a taxonomy that defines six types of deception:  perturbation, moving target defense, obfuscation, mixing, honey-x, and attacker engagement. These types are delineated by their \textcolor{black}{information structures}, agents, actions, and duration: precisely concepts captured by game theory. Our aims are to rigorously define types of defensive deception, to capture a snapshot of the state of the literature, to provide a menu of models which can be used for applied research, and to identify promising areas for future work. Our taxonomy provides a systematic foundation for understanding different types of defensive deception commonly encountered in cybersecurity and privacy.
\end{abstract}

%
%


%
%




\begin{bottomstuff}
This work is partially supported by an NSF IGERT grant through the Center for Interdisciplinary Studies in Security and Privacy (CRISSP) at New York University, by the grant CNS-1544782, EFRI-1441140, and SES-1541164 from National Science Foundation (NSF) and DE-NE0008571 from the Department of Energy. 
Research was also sponsored by the Army Research Laboratory and was accomplished under Cooperative Agreement Number W911NF-17-2-0104. The views and conclusions contained in this document are those of the authors and should not be interpreted as representing the official policies, either expressed or implied, of the Army Research Laboratory or the U.S. Government. The U.S. Government is authorized to reproduce and distribute reprints for Government purposes notwithstanding any copyright notation herein. 
Finally, the authors thank visiting student Richard Minicus for his contributions to the literature survey.

Author's addresses: Jeffrey Pawlick, Department of Electrical and Computer Engineering, New York University Tandon School of Engineering, 5 MetroTech Center, Brooklyn, NY, USA, and US Army Research Laboratory, 2800 Powder Mill Road, Adelphi, MD, USA, email: jpawlick@nyu.edu. Edward Colbert, US Army Research Laboratory, and Virgina Tech Intelligent Systems Laboratory, Hume Center for National Security and Technology, 900 N. Glebe Road, Arlington, VA, USA, email: ecolbert@vt.edu. Quanyan Zhu, Department of Electrical and Computer Engineering, New York University Tandon School of Engineering, email: quanyan.zhu@nyu.edu. 
\end{bottomstuff}

\maketitle

\section{Introduction}

\label{sec:Introduction}
\begin{quote}
\textquotedblleft All warfare is based on deception. Hence, when we
are able to attack, we must seem unable; when using our forces, we
must appear inactive; when we are near, we must make the enemy believe
we are far away; when far away, we must make him believe we are near.\textquotedblright ---Sun
Tzu, \emph{The Art of War}
\end{quote}
Deception has played an important role in the history of military combat. More generally, deception is commonplace in adversarial or strategic interactions in which one party possesses information unknown to the other. In this paper, we are motivated by recent  deception research in psychology, criminology, economics, and behavioral sciences.

\subsection{Deception Across Disciplines}

\label{sub:Deception-Across-Disciplines}

\subsubsection{\textcolor{black}{Military Applications}}

\textcolor{black}{Deception and secrecy demanded significant attention during World War II and the Cold War \cite{whaley2016practise,bell2017cheating}. But increasing globalization and the proliferation of communication technologies have recently created further challenges for mitigating deception. Increasing availability of information has led not only to more knowledge but also to worse confusion \cite{godson2011strategic}. National defense requires detailed study of military-relevant deceptions such as advanced persistent threats carried out by state actors \cite{bodmer2012reverse}.}

\subsubsection{Psychology and Criminology}

Research in psychology and criminology suggests that humans have poor
abilities to detect deception \cite{bond2008individual,vrij2008increasing}. 
One approach to address this shortcoming 
focuses on interview techniques. It has been shown that detection rates can be improved by tools that increase cognitive load
by asking suspects to recall events in reverse order, to maintain
eye contact, or to answer unexpected questions. Some of these have been incorporated into the investigative 
protocol known as the \emph{Cognitive Interview
for Suspects} (CIS) \cite{geiselman2012cognitive}. 
A second approach uses physiological indicators. For instance,
the \emph{Guilty Knowledge Test} (GKT) prompts a suspect with a list
of items---for example, a set of articles found at the scene of a
crime---and measures the suspect's physiological responses to each
item \cite{lykken1959gsr}. Signs of arousal in a suspect suggest that the suspect possesses guilty knowledge, because the articles are irrelevant to an innocent person.

\subsubsection{Cybersecurity}

Deception is used in cyberspace interactions for both defense and
attack. Examples of malicious deception include email phishing, man-in-the-middle attacks, and deployment 
of sybil nodes (nodes with forged identities) in a social network. Cybersecurity professionals have 
developed various defenses against these deceptions. For instance,
the field of \emph{adversarial machine learning} protects data analysis
algorithms against injection or manipulation of data by attackers
\cite{zhang2015secure,zhang2017game}, and the field of 
\emph{trust management} evaluates whether distributed nodes should
trust possibly-malicious signals \cite{pawlick2015flip,pawlick2017strategic}.
Examples of defensive deception include moving target
defense \cite{zhu2013game}, honeypot deployment \cite{Carroll2011}, and the use of mix networks \cite{zhang2010gpath}.

\subsubsection{Privacy Advocacy}

Recently, privacy advocates have designed technologies for Internet
users to \emph{obfuscate }the trails of their digital activity against
ubiquitous tracking. Privacy advocates
argue that developments such as third-party tracking
and persistent cookies have not been sufficiently regulated by law.
Therefore, there is a need for user-side technologies to provide proactive
privacy protection. 
One example is \emph{TrackMeNot}, a browser extension that periodically
issues random search queries in order to undermine search engine tracking
algorithms \cite{Howe2009}. Another example is \emph{CacheCloak},
which protects location privacy by retrieving location-based services
on multiple possible paths of a user. An adversary tracking the requests
is not able to infer the actual user location \cite{meyerowitzhiding2009}.
These are instances of deception that is designed
for benign purposes.

\subsubsection{Behavioral Economics}

In economics, the area of \emph{strategic communication} quantifies
the amount of information that can be transmitted between two parties
when communication is unverifiable 
\cite{crawford1982strategic,kartik2009strategic}. Communication can
be evaluated both strategically and physiologically. One recent paper
analyzes patterns of eye movement and pupil dilation during strategic
deception \cite{astyk2010pinocchio}. 
At the same time, research in behavioral economics finds that sometimes
economic agents choose not to deceive, even when it is incentive-compatible
\cite{gneezy2005deception,hurkens2009would,fischbacher2013lies}.
Subjects that exhibit so-called \emph{lying aversion} choose to maximize
their payoffs less frequently when it requires a lie than when it
requires a simple choice. This points towards the influences of morality and ethics 
on deception.

\subsubsection{Economic Markets}

In broader economics literature, \cite{akerlof2015phishing} argues that many markets
admit equilibria in which deceivers exploit those who are vulnerable. Akerlof and Shiller describe these interactions
in politics, pharmaceuticals, finance, and advertising. They use email
phishing as an analogy for any kind of deception in which an informed
``phisherman'' exploits the lack of knowledge or the psychological
vulnerabilities of a group of ``phools.'' The essential insight
is that opportunities for deception will be exploited in equilibrium. Across all six disciplines, deception involves interaction, conflict,
rationality, and uncertainty. 

\subsection{Cybersecurity and Privacy}

Major cybersecurity incidents in the past ten years include breaches of Home Depot in 2014, insurance company Anthem Inc,
in February of 2015, and the US Office of Personnel Management later
that year. The ten years from 2005 to 2015 featured over 4000 (publicized)
data breaches \cite{edwards2016hype}. During the same ten years, cyber attacks also affected physical
space. The power grid in Ukraine, Iranian nuclear centrifuges, and
an American water dam twenty miles north of New York were all infiltrated.
Since then, new attack vectors such as the Mirai botnet have highlighted
the potential to turn Internet of things (IoT) devices into domestic
cyber weapons. At the same time, IoT devices have raised new privacy concerns. Smartphones and
wearable electronics combine to collect sensitive data which can predict
\textquotedblleft a user\textquoteright s mood; stress levels; personality
type; bipolar disorder; demographics\textquotedblright{} \cite{peppetregulating2014}.
Advocates have responded by designing innovative privacy enhancing
technologies (PETs) and trying to formalize and defend appropriate
definitions of privacy \cite{nissenbaum2004privacy}.

\subsection{Defensive Deception}

Firewalls, cryptography, and role-based
access control, although essential components of any security strategy,
are unable to fully address these new cybersecurity and privacy threats. Adversaries often gain undetected, insider access to network
systems. They obtain information about networks
through reconnaissance, while defenders lack an understanding of the
threats that they face. 

Deception is crucial to counteract this information asymmetry. Consider
the following definition \cite{stanford2016deception}:
\begin{quote}
$\text{To deceive}\stackrel{\text{def}}{=}$ to intentionally cause
another person to acquire or continue to have a false belief, or to
be prevented from acquiring or cease to have a true belief.
\end{quote}
This definition is broad, which matches the diversity of
applications of defensive deception to cybersecurity and privacy.
On the other hand, the breadth of the term deception limits its depth.
Finer-resolution definitions are necessary to design deception techniques
for specific purposes. 

While affirming the breadth and variety of types of deception in our
area, we aim to tease out the distinctions between these types. To
some extent, this is already present in the literature; authors use
specific terms such as moving target defense, perturbation, and obfuscation
to refer to different types of deception. Nevertheless, the meaning
of these terms is not completely clear. For instance, what is the
difference between perturbation and obfuscation? \emph{We attempt
to answer this and other questions by proposing a taxonomy of defensive
deception for cybersecurity and privacy}.

\subsection{Game-Theoretic Taxonomy}

We are interested in understanding the question \emph{``What
are the various types of deception?''} from the viewpoint of quantitative
science. Since deceptive interactions are strategic confrontations
between rational agents, an appropriate quantitative science is \emph{game
theory}. Game theory 
\textcolor{black}{``can be defined as the study of mathematical models of conflict and cooperation between intelligent rational decision-makers''  \cite{myerson1991gametheory}.} Each player in a game makes decisions that may influence the welfare of the other players. Game theory applies
naturally to strategic and adversarial interactions in cybersecurity and privacy\footnote{\textcolor{black}{All of the works surveyed in this paper use ``non-cooperative game theory.'' In non-cooperative games, ``the players' choices are based only on their perceived self-interest, in contrast to the theory of cooperative games'' in which groups of players make decisions together in coalitions.  Still, ``non-cooperative players, motivated solely by self-interest, can exhibit `cooperative' behavior in some settings'' \cite{fudenberg1991game}.}}. 
\textcolor{black}{It  has  been  used  in  a  variety  of  cybersecurity  contexts. A few application areas include intrusion detection systems \cite{Alpcan2003}, adversarial machine learning \cite{zhang2015secure}, and communications jamming \cite{basar1983gaussian}. Applications in physical security include the ARMOR system for Los Angeles airport security \cite{pita2008deployed} and the PROTECT system for patrol routes in the port of Boston \cite{shieh2012protect}.}

Game theory offers a quantitative framework that can model the information structure, 
actors, actions, and duration of the various types of defensive deception.
As a systems science, it models the essential, transferable,
and universal aspects of defensive deception.
In doing so, it attempts to build up a science of security. 

Of course, game-theoretic models are useful only insofar as they capture
real features of cyberspace phenomena. Therefore, we need to apply
appropriate models to each type of deception. \emph{This is precisely
the motivation to develop a game-theoretic taxonomy of deception}:\emph{
}to identify the features of the various types of deception that must be captured
by game-theoretic models.

\subsection{Contributions and Related Work}

\label{sub:Contributions}

\begin{figure*}
\begin{centering}
\includegraphics[width=1\textwidth]{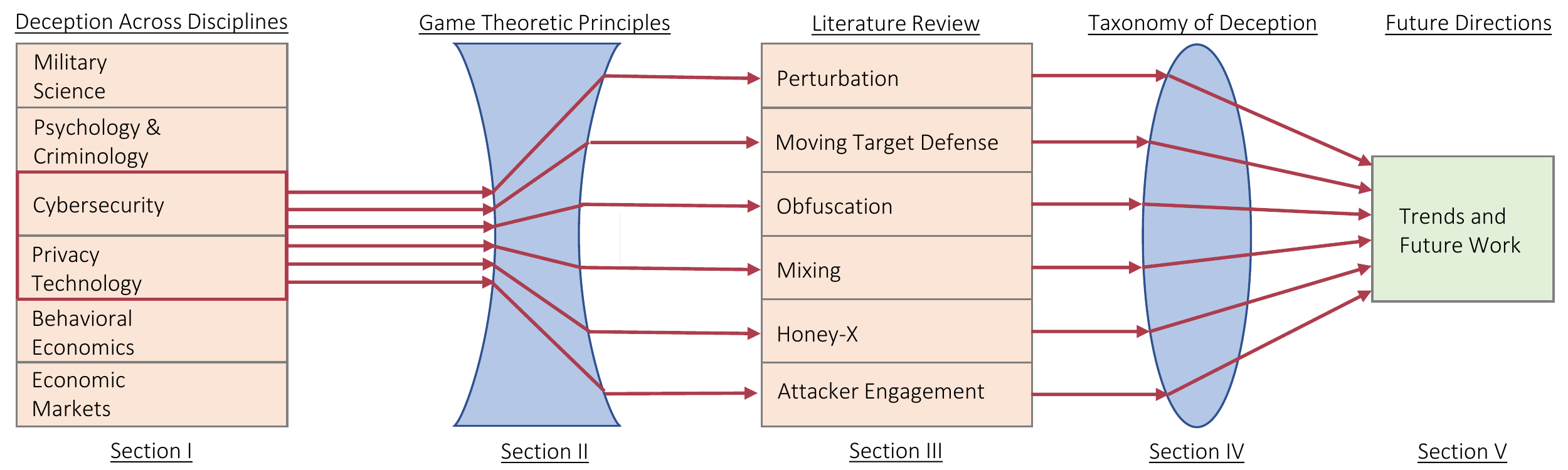}
\par\end{centering}

\caption{\label{fig:paperPlan}We begin by discussing deception across several
different disciplines, and then focus on defensive deception in cybersecurity
and privacy. Then we describe a set of game-theoretic principles which
are utilized in literature in this area, and  review the literature.
Finally, we create a taxonomy which defines each type of deception,
and we identify areas for future work.}
\end{figure*}

Figure \ref{fig:paperPlan} gives a conceptual outline of
the paper. We present the following principle contributions:
\begin{enumerate}
\item We review the game-theoretic models most commonly used to study cybersecurity
and privacy (Section \ref{sec:Review-of-Games}).
\item We survey the contributions of 24 recent (2008--2018) articles to
the field of game-theoretic models of defensive deception (Section
\ref{sec:Literature-Survey}).
\item We develop a taxonomy that defines specific types of
deception, including moving target defense, perturbation, mixing,
obfuscation, honey-x, and attacker engagement. We also show how the
game-theoretic concepts of private information, actors, actions, and duration
capture the essential differences between types of deception (Section
\ref{sec:Taxonomy}).
\item We discuss promising areas for future research. These include mimetic
deception, theoretical advances, practical implementations, and interdisciplinary
security (Section \ref{sec:Future-Directions}).
\end{enumerate}

All of the literature that we review 1) studies cybersecurity or privacy, 2) employs \emph{defensive} deception, and 3) uses game theory. This strictly constrains the scope of the paper. Notably, we have excluded papers that use deception for physical security (\emph{e.g.}, \cite{pita2008deployed,shieh2012protect},
\emph{etc.}). We have also excluded studies of malicious deception and the defensive methods that aim to mitigate it. 

\textcolor{black}{Surveys of game theory and cybersecurity can be found in \cite{roy2010survey,manshaei2013game,do2017game},
but these do not focus on deception in particular. 
Deception in the context of military applications has been well studied by Barton Whaley and coauthors \cite{bell2017cheating,rothstein2013art,whaley2016practise}. Some of these works include taxonomies. Specifically, \cite{rothstein2013art} makes a distinction between ``hiding the real'' and ``showing the false'' which we build upon in our taxonomy. Scott Gerwehr and Russell Glenn add that some deceivers aim to inject noise: randomness or intense activity that slows an adversary's ability to act. We find analogous aims in perturbation or obfuscation for the purposes of privacy. They also distinguish between static and dynamic deception. We adopt this distinction, since it motivates modeling efforts using one-shot or multiple-interaction games \cite{bennett2007counterdeception}.}

\begin{table}%
\tbl{Players, Types, Actions, and Utility Functions
for Three Games\label{tab:summaryGames}}{%
\centering{}{\small{}}%
\begin{tabular}{|c|c|c|c|c|}
\hline 
{\small{}Players $\mathcal{P}$} & {\small{}Types $\Theta$} & {\small{}Actions $\mathcal{A}$ } & {\small{}Utility $\mathcal{U}$ } & {\small{}Duration $\mathcal{T}$ }\tabularnewline
\hline 
\hline 
{\small{}Stackelberg game} & {\small{}Typically } & {\small{}$L:$ $a_{L}\in\mathcal{A}_{L}$} & {\small{}$L:$ $U_{L}(a_{L},a_{F})$} & {\small{}One-shot }\tabularnewline
{\small{}between leader $L$ } & {\small{}uniform} & {\small{}$F:$ $a_{F}\in\mathcal{A}_{F}$} & {\small{}$F:$ $U_{F}(a_{L},a_{F})$} & {\small{}leader-follower }\tabularnewline
{\small{}and follower $F$} &  &  &  & {\small{}structure}\tabularnewline
\hline 
{\small{}Nash game} & {\small{}Typically } & {\small{}$V:$ $a_{V}\in\mathcal{A}_{V}$} & {\small{}$V:$ $U_{V}(a_{V},a_{W})$} & {\small{}Simultaneous }\tabularnewline
{\small{}between semetric } & {\small{}uniform} & {\small{}$W:$ $a_{W}\in\mathcal{A}_{W}$} & {\small{}$W:$ $U_{W}(a_{V},a_{W})$} & {\small{}move structure}\tabularnewline
{\small{}players $V$ and $W$ } &  &  &  & \tabularnewline
\hline 
{\small{}Signaling game} & {\small{}$S$ has } & {\small{}$S:$ $a_{S}\in\mathcal{A}_{S}$} & {\small{}$S$ of each type } & {\small{}One-shot }\tabularnewline
{\small{}between sender $S$ } & {\small{}multiple} & {\small{}$R:$ $a_{R}\in\mathcal{A}_{R}$} & {\small{}$\theta\in\Theta:$ $U_{S}^{\theta}(a_{S},a_{R})$ } & {\small{}sender-receiver }\tabularnewline
{\small{}and receiver $R$} & {\small{}types $\theta\in\Theta$} &  & {\small{}$R:$ $U_{R}(\theta,a_{S},a_{R})$} & {\small{}structure}\tabularnewline
\hline 
\end{tabular}{\small \par}}
\end{table}%

\textcolor{black}{Our work is also related to \cite{rowe2016introduction}, which categorizes methods of deception according to techniques such as impersonation, delays, fakes, camouflage, false excuses, and social engineering. Finally, the present taxonomy can be compared and contrasted to those developed by Kristin Heckman et \emph{al}. \citeyear{heckman2015cyber}. One of these taxonomies breaks down malicious deception by stages including design of a cover story, planning, execution, monitoring, \emph{etc.} Another distinguishes between defensive deceptions based on whether they enable defensive visibility or restrict attacker visibility\footnote{\textcolor{black}{Several other works contribute to the effort to create a taxonomy of deception. Michael Handel makes a distinction between deceptions targeted against capabilities and those against intentions \cite{bennett2007counterdeception}, which we do not explore in the present work. Oltramari and coauthors  define an ontology of cybersecurity based on human factors \citeyear{oltramari2014building}, but this is more focused on trust than deception. Finally, Neil Rowe \citeyear{rowe2006taxonomy} 
describes a taxonomy of deception in cyberspace based on linguistic theory.}}. To the best of our knowledge, however, none of the related works develops a taxonomy based on game-theoretic distinctions. \emph{Our taxonomy distinguishes between different types of deception precisely in order to enable accurate game-theoretic modeling of each type.}}

\section{Review of Game-Theoretic Models}

\label{sec:Review-of-Games}

In this section, we introduce some of the most common game-theoretic models used in defensive deception for cybersecurity and
privacy: Stackelberg, Nash, and signaling games. Table \ref{tab:summaryGames} summarizes the components
of the games\footnote{All three games are static games (where we take one-shot interactions
to be static although they are not simultaneous). In addition, while
Stackelberg and Nash games typically do not include multiple types,
Bayesian Stackelberg and Bayesian Nash games do allow multiple types. }. These components define the structure of the taxonomy that
we develop in Section \ref{sec:Taxonomy}.

\subsection{Stackelberg Game}

\begin{figure}
\centering{}\includegraphics[width=0.45\columnwidth]{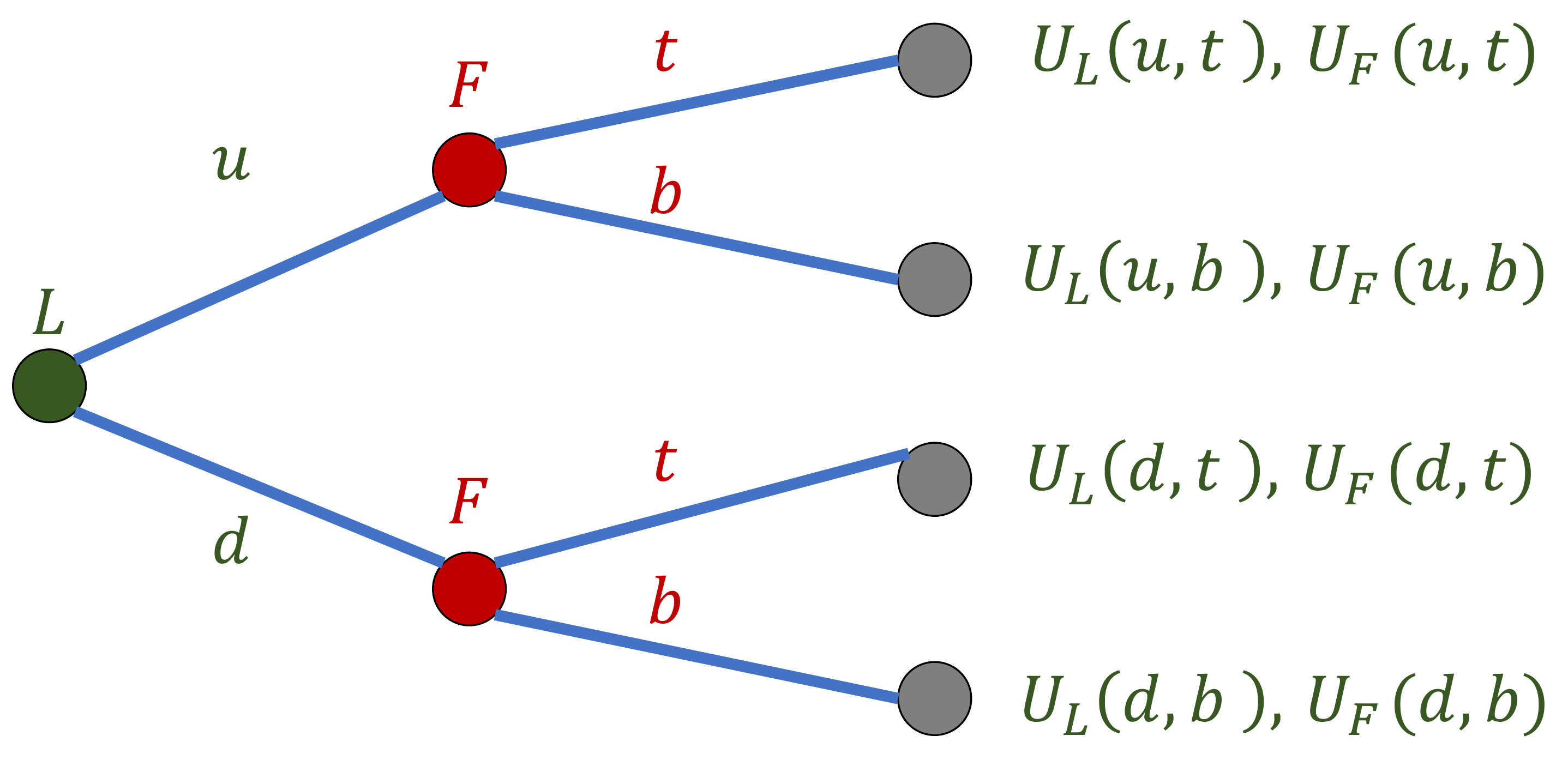}\caption{\label{fig:stack}Stackelberg games consist of a leader $L$ and a
follower $F.$ $L$ chooses an action $a_{L},$ and $F$ chooses a
best response $BR_{F}(a_{L}).$ $L$ takes this best response into
account when choosing $a_{L}.$ Stackelberg cybersecurity models often
consider the defender to be a leader, and the attacker to be a follower,
based on the assumption that the attacker will observe and react to
strategies chosen by the defender.}
\end{figure}
Stackelberg games \citeyear{vonstackelbergmarktform1934} are perhaps
the most fundamental game-theoretic interactions. They are characterized
by the following details.
\begin{enumerate}
\item \textbf{Players}: $\mathcal{P}=\left\{ L,F\right\} ,$ where $L$
is a \emph{leader} and $F$ is a \emph{follower}. 
\item \textbf{Actions}: The actions for player $L$ are given by $a_{L}\in\mathcal{A}_{L}.$
Figure \ref{fig:stack} shows a $2\times2$ game in which $\mathcal{A}=\{u,d\},$
and $u$ denotes moving \emph{up}, while $d$ denotes moving \emph{down.
}Player $F$ has actions $a_{F}\in\mathcal{A}_{F}=\left\{ t,b\right\} ,$
where $t$ denotes \emph{top} and $b$ denotes \emph{bottom}.
\item \textbf{Utilities}: After both players move, $L$ receives utility
$U_{L}\left(a_{L},a_{F}\right),$ and $F$ receives utility $U_{F}\left(a_{L},a_{F}\right).$ 
\end{enumerate}
In Stackelberg games, the follower moves \emph{after observing the leader's
action}. Often, cybersecurity models take the defender as $L$ and
the attacker as $F,$ assuming that the attacker will observe and
react to defensive strategies.

Stackelberg games are solved backwards in time. Let $\mathcal{P}(\mathcal{S})$
denote the power set of the set $\mathcal{S}.$ Then let $BR_{F}:\,\mathcal{A}_{L}\to\mathcal{P}(\mathcal{A}_{F})$
define a \emph{best response function} of the follower to the leader's
action. $BR_{F}\left(a_{L}\right)$ gives the optimal $a_{F}$ to
respond to $a_{L}.$ The best response function could also include
a set of equally good actions. This is the reason for the power set.
The best response function is defined by 
\begin{equation}
BR_{F}\left(a_{L}\right)=\underset{a_{F}\in\mathcal{A}_{F}}{\arg\max}\,U_{F}\left(a_{L},a_{F}\right).\label{eq:BR_F}
\end{equation}
Based on anticipating $F$'s best response, $L$ chooses optimal action
$a_{L}^{*}$ which satisfies
\begin{equation}
a_{L}^{*}\in\underset{a_{L}\in\mathcal{A}_{L}}{\arg\max}\,U_{L}\left(a_{L},BR_{F}\left(a_{L}\right)\right).
\end{equation}
Then, in equilibrium, the players' actions are $\left(a_{L}^{*},a_{F}^{*}\right),$
where $a_{F}^{*}\in BR_{F}\left(a_{L}^{*}\right).$

\subsection{Nash Game}

\begin{figure}
\centering{}\includegraphics[width=0.45\columnwidth]{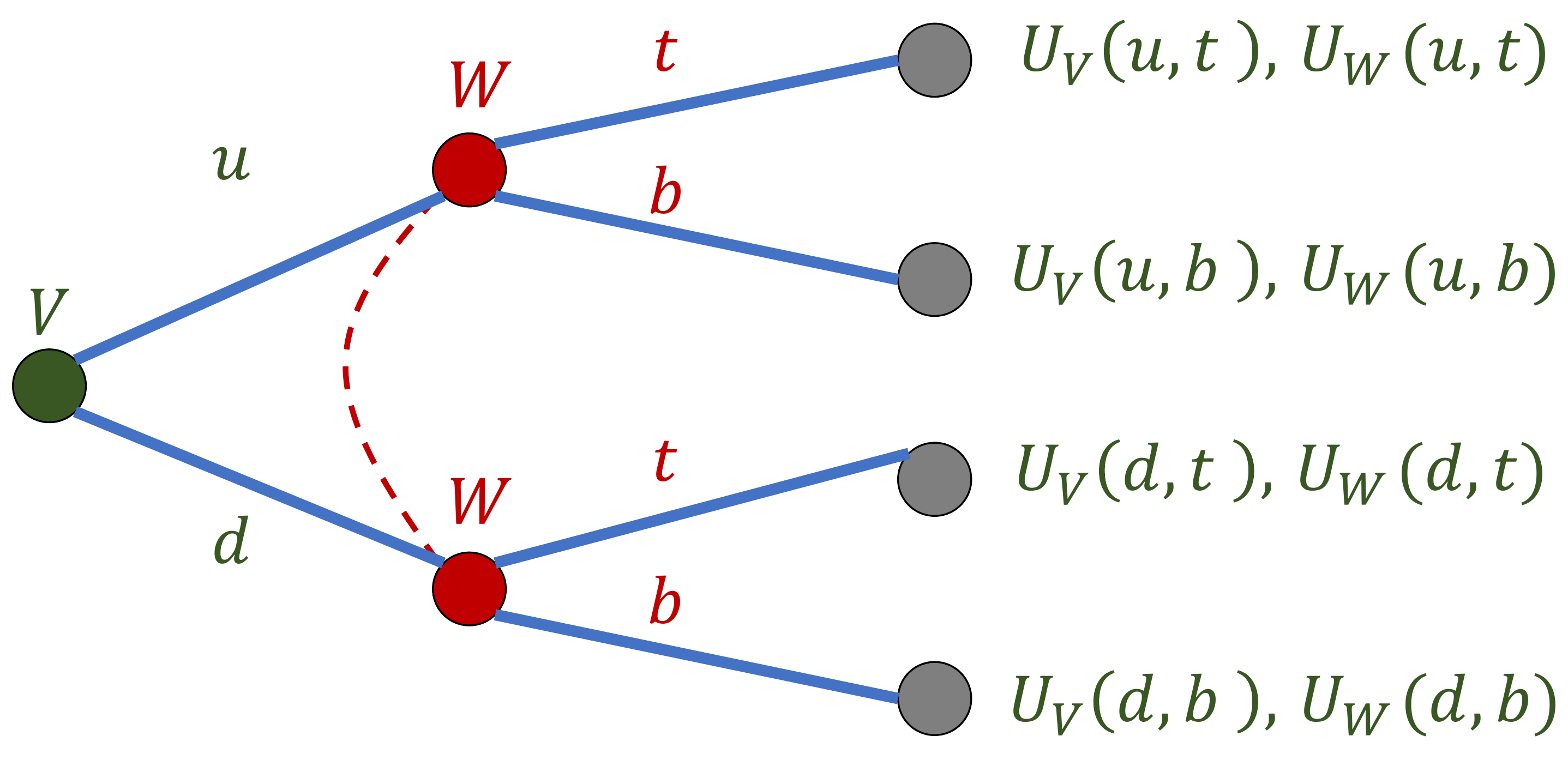}\caption{\label{fig:nash}Nash games are interactions with prior commitment.
In this case, the dashed line indicates that $W$ does not know which
node describes the game, because she does not know which move $V$
has chosen.}
\end{figure}
While in Stackelberg games players move at different times, in Nash
games \citeyear{nash1950equilibrium} players move simultaneously. More
precisely, Nash games are games of \emph{prior commitment}, in which
each player commits to his or her strategy before knowing the other
player's move. Typically, two-player games of prior commitment are
shown in matrix form. Figure \ref{fig:nash}, however, gives a tree
diagram of a two-player game in order to show the difference between
this game and a Stackelberg game. Players $V$ and $W$ act simultaneously,
or at least without knowing the other player's action. The dashed
line connecting the two nodes for $W$ denotes that $W$ does not
know which node the game has reached, because she does not know which
move $V$ has chosen.

The concept of \emph{Nash equilibrium} requires each player to choose
a strategy which is optimal given the other player's strategy. Let
$BR_{V}:\,\mathcal{A}_{W}\to\mathcal{P}(\mathcal{A}_{V})$ be defined
such that $BR_{V}\left(a_{W}\right)$ gives the set of actions for
$V$ which optimally respond to $W$'s action $a_{W}.$ Let $BR_{W}$
be defined similarly. Then a pure strategy Nash equilibrium \citeyear{nash1950equilibrium}
is given by a pair $\left(a_{V}^{*},a_{W}^{*}\right)$ such that
\begin{equation}
a_{V}^{*}\in BR_{V}\left(a_{W}^{*}\right),
\end{equation}
\begin{equation}
a_{W}^{*}\in BR_{W}\left(a_{V}^{*}\right).
\end{equation}
Nash equilibrium often requires players to choose actions according
to probability distributions. These strategies are called \emph{mixed
strategies}. Mixed strategies implement the basic idea of randomizing
allocations of defense assets in order to avoid leaving vulnerabilities
open to an attacker.

\subsection{Signaling Game}

\begin{figure*}
\centering{}\includegraphics[width=0.85\textwidth]{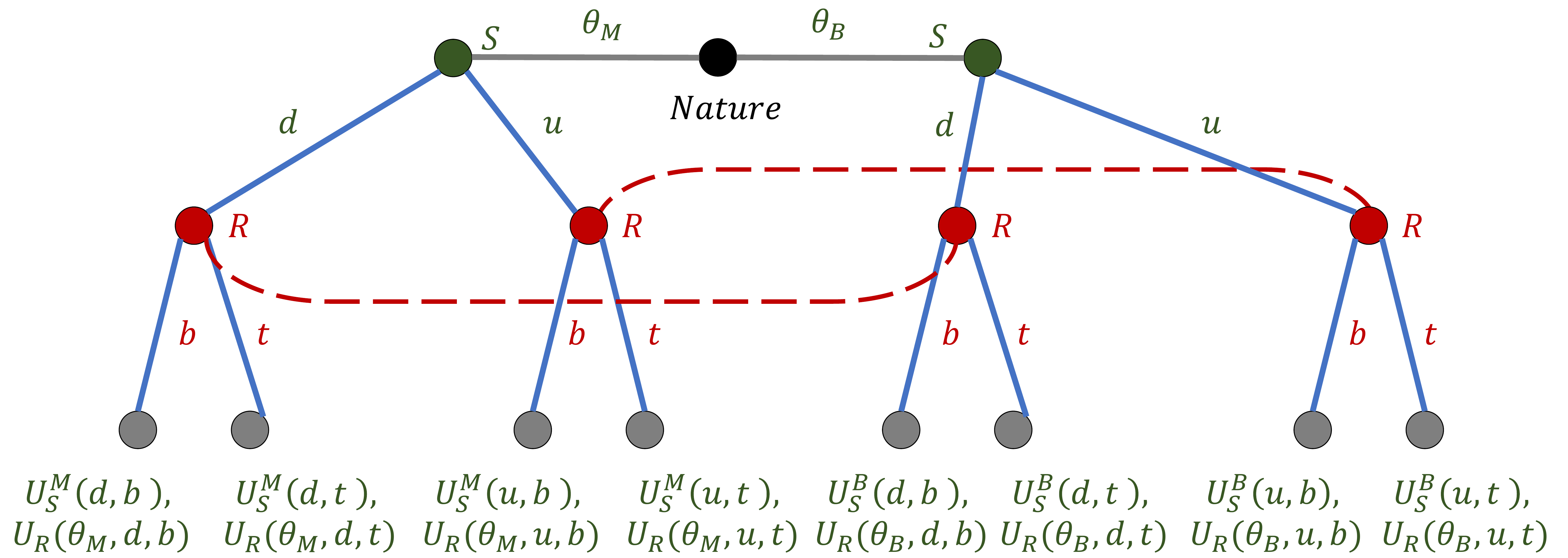}\caption{\label{fig:signaling}The figure depicts a signaling game in which
a sender $S,$ who has access to private information, transmits a
message to a receiver $R.$ The message is not verifiable, so $R$
does not know the underlying information with certainty. In separating
and partially-separating equilibria, however, it is incentive-compatible
for $S$ to transmit a message that at least partially reveals his
private information.}
\end{figure*}
Signaling games, like Stackelberg games, are two-player dynamic interactions
(Fig. \ref{fig:signaling}). Signaling games typically label the players
as \emph{sender} $S$ and \emph{receiver} $R.$ The sender has access to some information
unknown to the receiver. This is called the sender's \emph{type} $\theta\in\Theta.$
The receiver only learns about the type based on the sender's action.
For this reason, the sender's action (here $a_{S}$) is referred to
as a \emph{message}. The message need not correspond to the sender's
type.

In this example, the set of types of $S$ is $\Theta=\{\theta_{B},\theta_{M}\},$
where $\theta_{B}$ represents a \emph{benign} sender, and $\theta_{M}$
represents a \emph{malicious }sender. Let $p(\theta)$ denote the
\emph{prior probability }with which $S$ has each type $\theta\in\Theta.$
The utility functions depend on type. $U_{S}^{M}\left(a_{S},a_{R}\right)$
and $U_{S}^{B}\left(a_{S},a_{R}\right)$ give the utility functions
for malicious and benign senders, respectively. $U_{R}\left(\theta,a_{S},a_{R}\right)$
gives the utility function for the receiver when the type of $S$
is $\theta,$ the sender message is $a_{S},$ and the follower action
is $a_{R}.$ $R$ forms a \emph{belief} $\gamma\left(\theta\,|\,a_{S}\right)$
that $S$ has type $\theta$ given that he sends message $a_{S}\in\mathcal{A}_{S}.$
To be consistent, this belief should be updated in accord with Bayes'
law. A strategy pair in which $S$ and $R$ maximize their own utilities,
together with a belief which is consistent, form a \emph{perfect Bayesian
Nash equilibrium }(PBNE) (\emph{c.f.} \cite{fudenberg1991game}).
In some PBNE, $S$ can cause $R$ to form a specific false belief.

\section{Literature Survey}

\label{sec:Literature-Survey}In this section, we survey existing
literature in game-theoretic approaches to defensive deception for
cybersecurity and privacy.   Tables  \ref{tab:justification1}-\ref{tab:justification2} 
in the appendix list the papers, which we now discuss from top to bottom.

\subsection{Perturbation}

\label{sub:Perturbation}
\begin{figure}
\begin{centering}
\includegraphics[width=0.7\columnwidth]{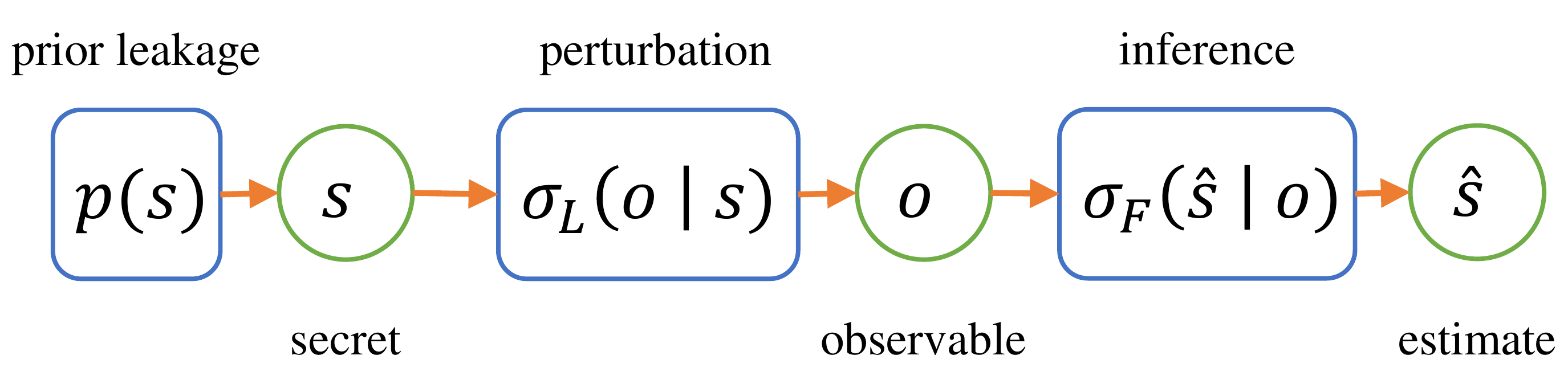}
\par\end{centering}

\caption{\label{fig:infoSharing}In the Information Sharing Framework from
Shorki \citeyear{shokri2015privacy}, a user hides a secret $s$ by releasing
an observable $o$ according to the channel probability $\sigma_{L}(o\,|\,s).$
The adversary uses inference $\sigma_{F}(\hat{s}\,|\,o)$ to obtain
estimate $\hat{s}.$ This is a Stackelberg game, because the adversary
knows the perturbation mechanism when he designs an inference attack. }
\end{figure}

The first set of papers study privacy that is obtained
by perturbing sensitive data. Chessa et \emph{al}. \citeyear{chessa2015game}
model the interaction between a set of users who contribute data in
order to identify mechanisms by which the learning agent can improve
the quality of its estimation. The users have two strategic variables:
whether to contribute at all, and the level of precision to use if
they do contribute. They face a trade-off between privacy and the
benefit of contributing to data analysis. \textcolor{black}{The paper takes this benefit---although possibly difficult to quantify---to
be a non-monetary interest in data as a public good.} Chessa et \emph{al}.
study both homogeneous and heterogeneous populations of users. The
interactions between these populations of users are modeled as Nash
games. After finding that the equilibrium in these games is worse
than the population-optimal behavior due to the selfishness of the
agents, the authors consider a mechanism for the data collector to
improve the equilibrium. They find that an analyst can shift the equilibrium
of the game towards more accurate estimation by imposing a minimum
level of precision in the strategy space of the users. The paper studies
computation of the average of a data set as a foundation for more
complex statistics \cite{chessa2015game}.

Shokri \citeyear{shokri2015privacy} also addresses the issue of data
privacy for users who send private information to an untrusted source.
 Shorki's work concerns two types of privacy, differential privacy
and distortion privacy. Differential privacy obtains a guaranteed
limit on the amount of information that users leak to an attacker,
but does not tell the user the actual level of privacy achieved. Distortion
privacy is a measure of the actual error of the attacker\textquoteright s
inferences based on the information that is leaked. The calculation
of this metric, however, requires that the attacker have prior knowledge
before making an observation. Using a non-zero-sum Stackelberg game,
Shokri develops a model for joint distortion-differential privacy.
Figure \ref{fig:infoSharing} depicts the interaction. The user leads
by choosing a protection mechanism. The attacker follows, designing
an inference attack and aiming to minimize user privacy. The inference
attack is the best response to the user\textquoteright s protection
mechanism. The author develops a linear program to obtain the equilibrium
\cite{shokri2015privacy}.

Alvim et \emph{al}. \citeyear{alvim2017leakage} study information privacy
within the setting of information theory. Specifically, their paper
formulates information transmission and leakage using quantitative
information flow. A sender possesses a secret, which a defender only
knows with some prior probability. The sender transmits the secret
through a channel, and the channel leaks some information to the adversary.
Using this information, the adversary forms an \emph{a posteriori}
belief about the secret. A quantity called the posterior vulnerability
of the secret quantifies the degree of the adversary's knowledge.
The quantification can be the entropy of the posterior distributions,
although a general convex function can also be used. Alvim et \emph{al.}
observe that the expected utility becomes convex, rather than linear,
in the mixed-strategy probabilities. For this reason, they call the
game an \emph{information leakage game} \cite{alvim2017leakage}.
 This formulation makes useful connections to the abundant tools
available in information theory. 

Theodorakopoulos et \emph{al}. \citeyear{theodorakopoulos2014prolonging}
address the issue of privacy for location-based services (LBS). The
authors emphasize that human locations are not found at discrete spatial-temporal
points, but rather follow a trajectory. Theodorakopoulos et \emph{al}.
propose user-centric location privacy preserving mechanisms (LPPMs) that send LBS pseudolocations to protect
not only the privacy of past and future locations, but also transitions
between locations and locations between LBS queries. As a Bayesian
Stackelberg game leader, the LPPM strategically sends a perturbed
pseudolocation to the LBS to protect its exact real location. Privacy
provided by the LPPM is measured by the error with which an adversary
deduces the real location of the target. Quality is determined by
how much the altered location data from the LPPM affects the functionality
of the LBS. Both privacy and quality values are computed for users
that prioritize 1) protecting past locations or 2) protecting future
locations. The authors' contribution is to improve upon \textquotedblleft trajectory-oblivious\textquotedblright{}
LPPMs. They evaluate the LPPM
using a set of mobility traces for taxi cabs in the San Francisco
Bay area \cite{theodorakopoulos2014prolonging}. 

Of course, a large body of literature studies the use of perturbation
to protect privacy (especially as quantified by differential privacy).
We have only focused on the subset of this literature that uses game
theory. We also note that perturbation is very similar to obfuscation,
which we survey in Subsection \ref{sub:Obfuscation}. Section \ref{sec:Taxonomy}
defines the difference between the two types and explains why we have
categorized some papers as perturbation but others as obfuscation.

\subsection{Moving Target Defense}

\label{sub:Moving-Target-Defense}
\begin{figure}
\begin{centering}
\includegraphics[width=0.45\columnwidth]{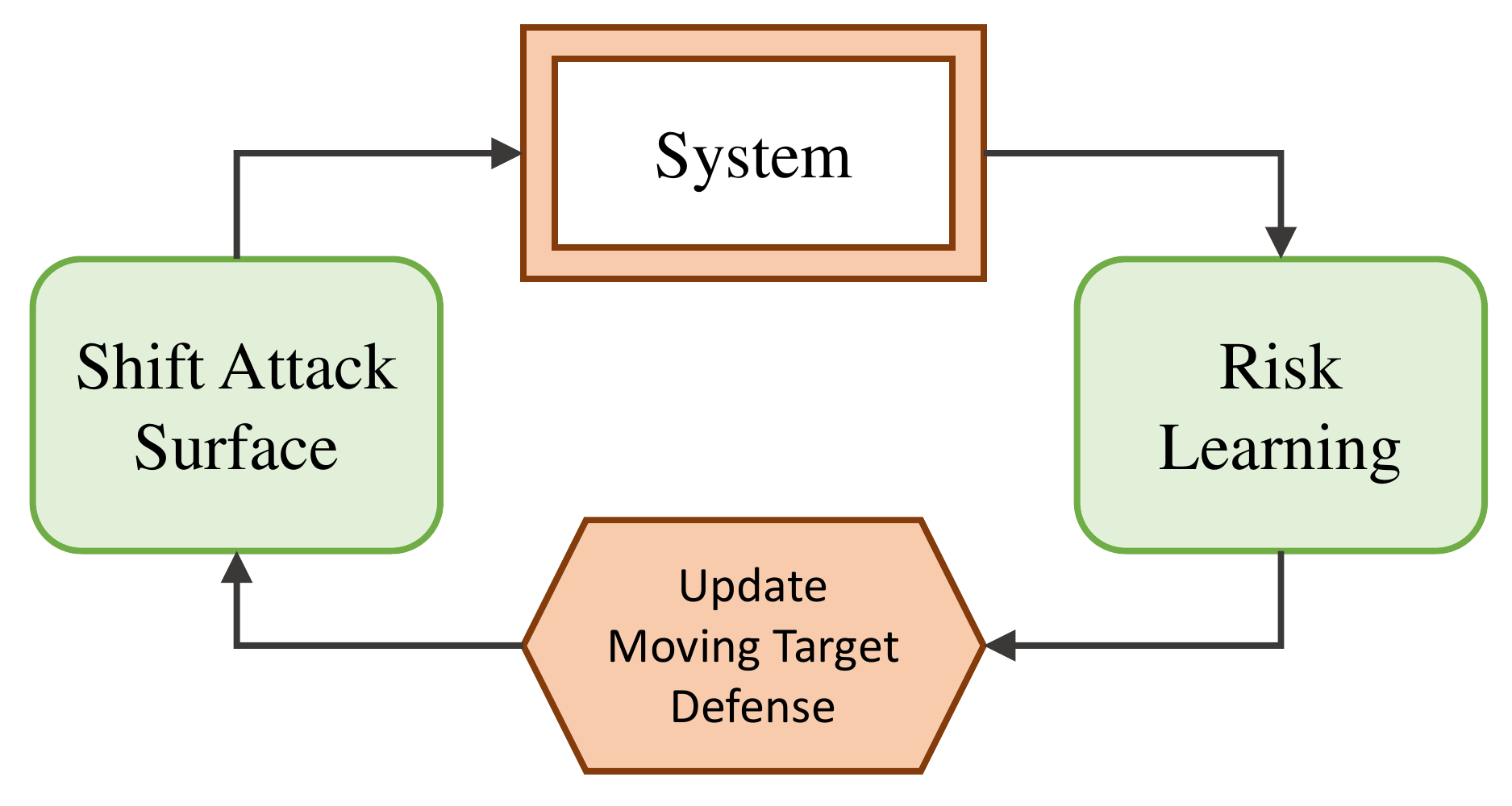}
\par\end{centering}

\caption{\label{fig:MTD}The moving target defense designed in \cite{zhu2013game}
updates the configuration of a system each round based on learning
the risk of attack that the system faces.}
\end{figure}
Next, we discuss papers that study moving target defense. Moving target defense 
invokes concepts of agility, changing attack surfaces, and random
configurations. Two different game-theoretic
concepts are often used to model moving target defense. They differ in whether time
is considered explicitly. One approach is to model randomness using
mixed strategies, with the implicit understanding that new realizations
from a random variable will be performed over time. A recent paper
by Rass et \emph{al. }notes a shortcoming of this approach: utility
functions based on mixed strategies do not inherently capture the
cost of switching from one pure strategy to another over time \citeyear{rass2017controlmixed}.
The second approach to moving target defense is to use Markov decision processes.
In Markov decision processes for moving target defense, the defense configuration is considered a state,
and switching costs can be modeled explicitly.


\textcolor{black}{Sengupta et al. \citeyear{sengupta2018moving} consider the placement of intrusion detection systems (IDS) in a cloud-based architecture. Two cloud servers and a cloud controller are connected to the Internet via a physical router. An attacker who may be located either outside or inside the network attempts to gain access. The network designer is able to place a limited number of IDS. The design problem is where to place them. In order to limit the knowledge of the attacker, these placements are performed probabilistically. Hence, this is a moving target defense. First the defender chooses a probability distribution for the locations of the IDS. Then the attacker observes this distribution and he chooses which system to attack. The interaction is modeled by a Stackelberg game with mixed strategies. The game can be solved by a mixed-integer linear program, although ultimately the most efficient formulation is as a mixed-integer quadratic program, using a branch-and-cut algorithm. One interesting contribution of this paper is that the authors include a concise taxonomy of moving target defense. They break moving target defense into four categories: shifting the exploration surface, detection surface, attack surface, or prevention surface. This could be useful to add a layer to the taxonomy proposed in the present paper.}

Zhu and Ba\c{s}ar \citeyear{zhu2013game} use Markov decision processes to model moving target defense  
for network security. An attacker and defender play a series of zero-sum
Nash games in which the defender chooses the arrangement of systems
with various vulnerabilities, and the attacker selects an attack path
that depends for its success on the arrangement of the vulnerabilities.
See Fig. \ref{fig:MTD}. Tools from control theory are used to analyze
the dynamics of the multiple-round interaction. Zhu and Ba\c{s}ar
show that the steady states of the learning dynamics correspond to
equilibrium points in the Nash games. The authors illustrate the game
and resulting dynamics with a numerical example. This work points
towards modeling the interaction between the attacker and defender
as a stochastic game in which the defender would attempt to minimize
the total risk over the whole game, rather than optimizing his risk
at each individual stage \cite{zhu2013game}.

Feng et \emph{al}. \citeyear{feng2017mtd} study moving target defense using a combination
of a Markov decision process and a Stackelberg game between an attacker and a defender
of an abstract resource such as a network asset. The defender
begins each turn by claiming a moving target defense strategy, which lasts for $\mathcal{T}$
periods. The defender controls a Markov decision process in which the states represent
security configurations of the system. Then the attacker follows by
choosing one state to attack. If the attack is successful, then the
attacker obtains a reward. The problem can be expressed in min-max
form, which the authors show can be reduced to simply a minimization
problem. They obtain an algorithm which has $O(|V|^{2}\log|V|)$ complexity,
where $|V|$ is the number of valid states. One insight from the paper
is that the defender's cost can be reduced more easily by increasing
the degree of the switching graph than by reducing the costs of switching
\cite{feng2017mtd}.


\subsection{Obfuscation}

\label{sub:Obfuscation}
\begin{figure}
\begin{centering}
\includegraphics[width=0.6\columnwidth]{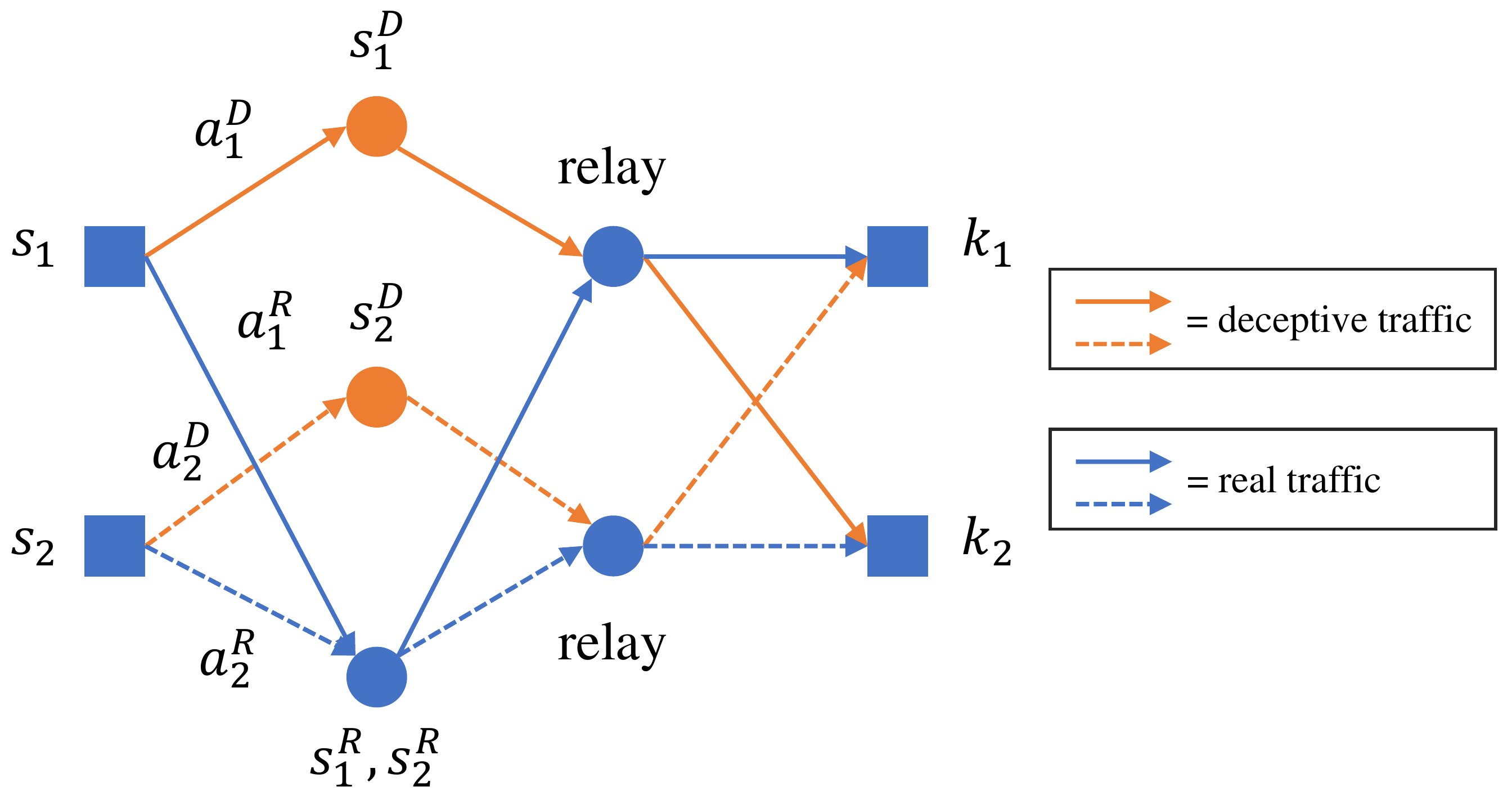}
\par\end{centering}
\caption{\label{fig:decepRout}Deceptive routing \cite{clark2012deceptive}
protects traffic routed from source $s_{1}$ to sink $k_{1}$ and
source $s_{2}$ to sink $k_{2}$ from jamming attacks. Player $1$
sends real traffic with volume $a_{1}^{R}$ to legitimate node $s_{1}^{R},$
but also sends decoy traffic with volume $a_{1}^{D}$ to deceptive
node $s_{1}^{D}.$ Player $2$ has a similar strategy. With some probability,
the adversary attacks the deceptive rather than the real traffic. }
\end{figure}
The third set of papers relates to hiding valuable information using
external noise. Clark et \emph{al}. \citeyear{clark2012deceptive} model
a network in which information is sent from a source to a destination
over multiple nodes. See Fig. \ref{fig:decepRout}. \textcolor{black}{The interaction is modeled using a Stackelberg game in which the defender leads by choosing the flow rates of deceptive and real traffic. The attacker observes these flows, and then follows by choosing a path to attack.} The authors consider situations in which sources do and do not cooperate to obfuscate
their traffic. They model the compounded pathway congestion that
comes with the addition of fake data, and the possibility that information
sources may take into account the delays being experienced by other
nodes in the network when determining their own data pathways and
flow rates. A simulation illustrates that altruistic sources improve
node functionality. Clark et \emph{al}. suggest that future work can
study other measurements of the efficacy of deception and an incomplete
information model in which nodes are only partially informed about
the attacker\textquoteright s cost and utility functions \citeyear{clark2012deceptive}. 

Zhu et \emph{al}. \citeyear{zhu2012deceptive} build upon this idea in
order to consider both 1) deceptive routing, and 2) strategic choices
of flow rates. In order to model both of these conflicts, the authors
formulate multiple-player Stackelberg games in which the defender
leads and the attackers follow. The routing game requires an extension
to Stackelberg games to allow $N+3$ players, where $N$ is the number
of attackers. Zhu et \emph{al}. quantify the effectiveness of the
mechanism using a metric called \emph{value of deception}, a ratio
of the utility achieved with and without the use of deception. Simulations
illustrate that defender utility decreases approximately asymptotically
in the attacker's budget, and increases approximately linearly in
the number of hops. Future work could include the introduction of
learning to similar algorithms, as well as the application of the 
model to multiple routing pathways \cite{zhu2012deceptive}.

While \cite{clark2012deceptive,zhu2012deceptive} apply obfuscation
to security, this type of deception also applies directly to privacy.
Two recent papers by Pawlick and Zhu \citeyear{pawlick2016stackelberg,pawlick2017meanfield}
model the long-run impact of obfuscation technologies such as \emph{TrackMeNot}
\cite{Howe2009} (a browser add-on which issues randomized search
engine queries to obfuscate a user\textquoteright s search profile)
and \emph{ScareMail} \cite{grosser2014scare} (which appends security-sensitive
words to the end of every email in an effort to undermine mass surveillance).
These tools add irrelevant data (\emph{e.g.}, randomized browser searches)
to relevant data (\emph{e.g}., real browser searches). The first
paper develops a Stackelberg game in which the leader is a machine
learning or tracking agent and the follower is a user. The learning agent has the option to promise a level
of differential privacy protection. The user reacts to this level,
and chooses whether to obfuscate the data himself. Under certain conditions,
it is incentive-compatible for the learning agent to promise some
level of privacy protection in order to avoid user obfuscation \cite{pawlick2016stackelberg}.
The second paper extends this scenario to multiple users through a
mean-field game model. In a mean-field game, each user responds to
the average obfuscation of the field of other users. This extended
model predicts a threshold of perturbation beyond which users will
adopt obfuscation in a cascading manner. The paper also identifies
conditions under which obfuscation does not motivate privacy protection
from the learner but only results in data pollution \cite{pawlick2017meanfield}. \textcolor{black}{These papers could benefit from future measurements of users' preferences in experimental or empirical settings.}

\subsection{Mixing}

\label{sub:Mixing}
\begin{figure}
\begin{centering}
\includegraphics[width=0.45\columnwidth]{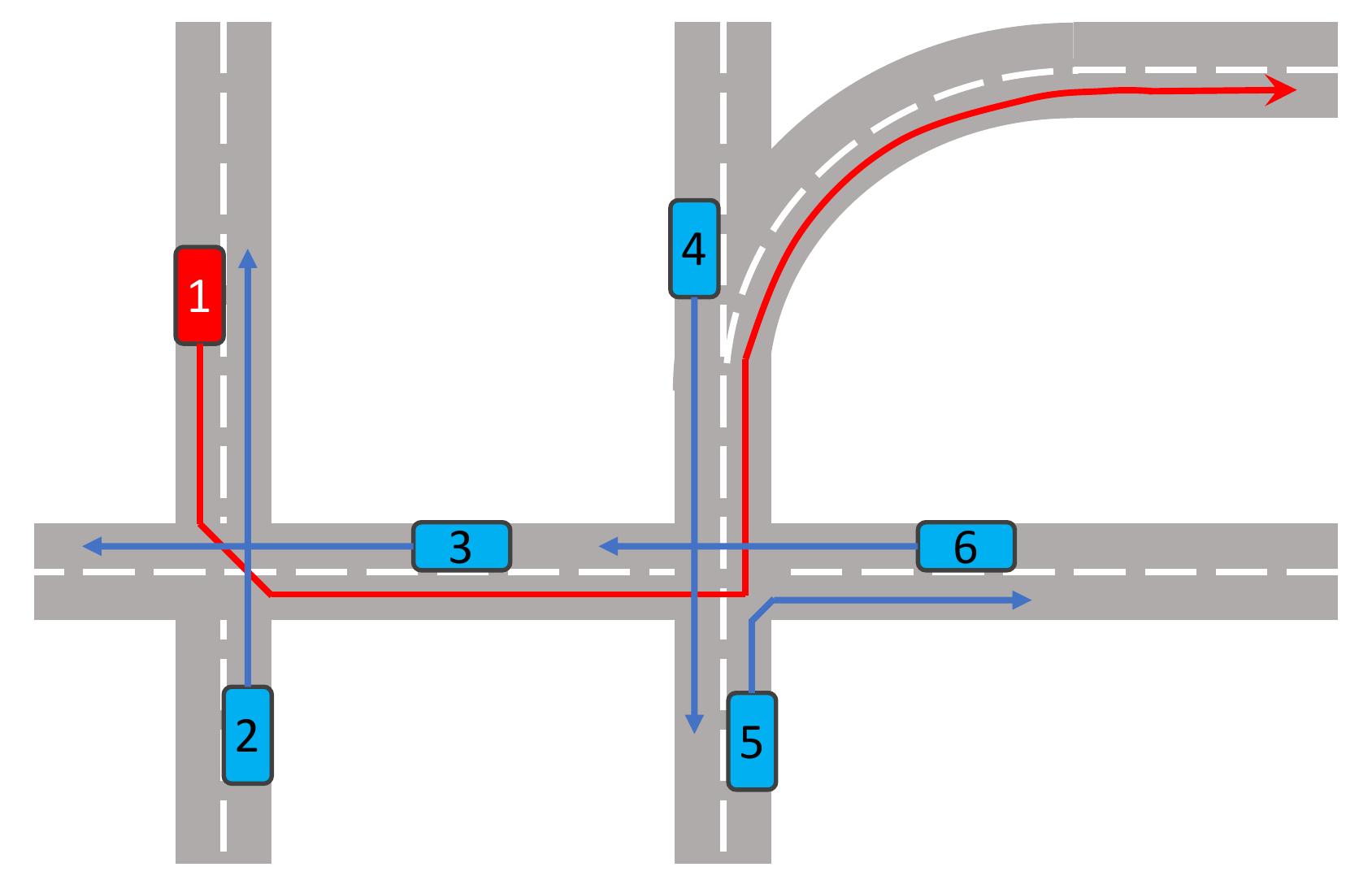}
\par\end{centering}

\caption{\label{fig:mixing}Freudiger et \emph{al}. \citeyear{freudiger2009loc}
study pseudonym swapping for location privacy in vehicle ad hoc networks.
As Vehicle $1$ turns left at its first intersection, it can swap
pseudonyms with Vehicle $2$ and Vehicle $3.$ At its second intersection,
it can swap with Vehicles $4,$ $5,$ and $6.$ If these other vehicles
have recently swapped, however, they have little incentive to swap
again. Thus, the interactions become multiple-player, simultaneous-move
games. }
\end{figure}
Techniques for both security and privacy use the idea of mixing in
order to prevent linkability. Zhang et \emph{al}. \citeyear{zhang2010gpath}
study anonymity of the \emph{Tor} network. When information is sent
over \emph{Tor}, it passes through three nodes before reaching its
destination, in an attempt to make the entry and exit nodes unlinkable.
 Zhang et \emph{al}. point out that since connectivity on \emph{Tor}
is volunteer-based, attackers could collude to link the entry and
exit nodes. They propose\emph{ gPath}, which operates on three principles:
1) an exit-eligible node should not be considered for the entry, 2)
the middle node should have equal or greater bandwidth than the entry
node, and 3) the exit node should be chosen based on the bandwidths
of the entry and middle nodes. The authors construct utility functions
for the user and attacker. The players use these functions to compute
best responses to each other's strategies in four rounds. This is
different from Nash equilibrium, since the strategies do not necessarily
converge to mutual best responses. Simulations performed on a snapshot
of the \emph{Tor} network from 2010 suggest that \emph{gPath} reduces
the probability of a successful attack \cite{zhang2010gpath}.

The idea of mixing can also be applied to cyber-physical systems such
as vehicle ad hoc networks (VANETs). Freudiger et \emph{al}. \citeyear{freudiger2009loc} use locations
called \emph{mix zones} for location privacy in VANETs. Vehicles in
these networks communicate using pseudonyms in order to achieve anonymity.
Nevertheless, a constant trace of communications from a single pseudonym
creates a link between a vehicle\textquoteright s origin and
destination. One technology proposed to overcome this linkability
is pseudonym swap in mix zones (Fig. \ref{fig:mixing}). At locations
such as intersections, multiple vehicles simultaneously exchange their
pseudonyms, so that the identities of vehicles that enter the swap zone
can only be linked with the identities of vehicles that exit the swap
zone with some probability. Freudiger et \emph{al}. 
note that this is a strategic interaction, because swapping involves
some cost. Selfish actors may decline to swap. Nash equilibrium is
used since pseudonym swapping is a simultaneous interaction. Freudiger
et \emph{al}. consider both complete and incomplete information games.
In the incomplete information games, each player is unaware of the
privacy loss which the other players have incurred since their last
pseudonym swap (and thus their willingness to swap again). Interestingly, VANET nodes 
\textcolor{black}{tend to swap pseudonyms more often when pseudonym swapping is expensive than when it is cheap \cite{freudiger2009loc}.}
One challenge is that swap locations are presumably those in which
communication is most necessary and should not be interrupted (such
as at an intersection). 

Lu et \emph{al}. \citeyear{lu2012pseudonym} address this dilemma by proposing
pseudonym changes at social spots, where many vehicles often stop
for a period of time. Small social spots, \emph{e.g.}, stoplights,
are places where vehicles stay briefly. Large social spots, such as
parking lots, are places where vehicles remain for longer duration.
A privacy mechanism---Key-insulated Pseudonym Self-Delegation (KPSD)---is
proposed. Authorized keys (pseudonyms) for each vehicle are kept in
a secure location by a Trusted Authority (TA), usually at the car
owner\textquoteright s home. The TA loads onto the OnBoard
Unit (OBU) a set of temporary
keys before each new trip. As in \cite{freudiger2009loc}, the adversary
is global (\emph{i.e.}, has full access to radio network safety messages
that provide vehicle time, location, speed, content and pseudonym)
and external (only listening in on conversations, not engaging in
vehicle sabotage), with access to the vehicle\textquoteright s OBU. This KPSD model is also unique compared to other models
because of its ability to prevent theft. Since pseudonyms are changed
at social spots and since the keys are not kept with the vehicle but
instead with the TA, a thief cannot generate temporary keys needed
for vehicle operation. Lu et \emph{al}. incorporate a game-theoretic
argument (based on Nash equilibrium) to discuss the scenarios under
which KPSD is incentive-compatible \cite{lu2012pseudonym}.

\subsection{Honey-X}

\label{sub:Honey-X}
\begin{figure}
\begin{centering}
\includegraphics[width=0.85\columnwidth]{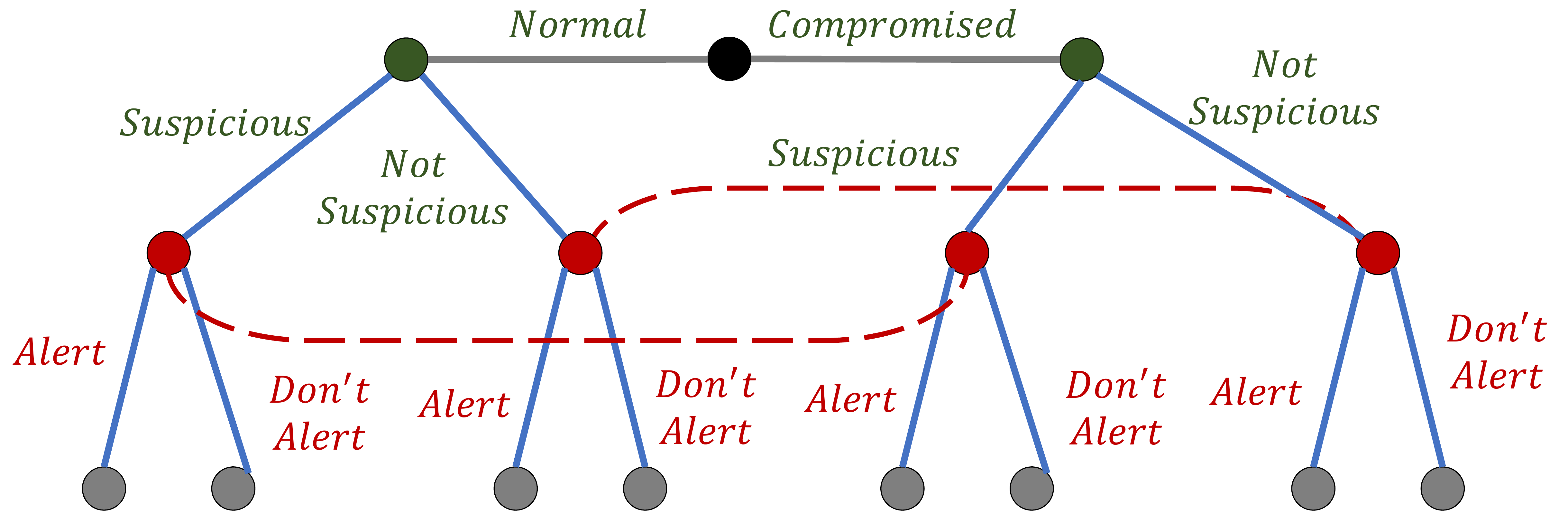}
\par\end{centering}

\caption{\label{fig:avatar}Mohammadi et \emph{al}. \citeyear{mohammadi2016game}
use signaling games to model the use of fake avatars as a type of
honey-x. The avatars observe suspicious or non-suspicious activity
from other user accounts in the network, and must form a belief about
whether the other users are legitimate accounts or compromised accounts.
The goal of the avatars is to correctly report compromised accounts,
while avoiding false positives. }
\end{figure}
By honey-x, we refer to deception which uses techniques named with
the prefix honey (\emph{e.g.}, honeypot, honeynet, honeybot, etc.).
\textcolor{black}{In many implementations of these techniques, defenders have a limited capacity to monitor honeypots, and attackers have constrained time and available technologies in order to detect honeypots. Therefore, game-theoretic trade-offs arise.} 
Carroll and Grosu \citeyear{Carroll2011} study a network in which a defender
can disguise honeypots as real systems (or reveal them as honeypots)
and real systems as honeypots (or reveal them as real systems). The
goal is to make attackers waste resources attacking honeypots and
avoiding real systems. In the proposed signaling game, the defender
is the sender and the attacker is the receiver. The equilibrium concept
is PBNE. \textcolor{black}{Because the defender never wants to reveal the system type to the attacker, the game does not support any
separating equilibria.} Carroll and Grosu analyze both pooling and
hybrid (partially separating) equilibria. Hybrid equilibria demonstrate
cases in which the defender has an incentive to partially reveal the
actual system type \cite{Carroll2011}. 

\textcolor{black}{This work by Carroll and Grosu has inspired several other papers. \c{C}eker \emph{et al.} \citeyear{ceker2016deception} build on this model for honeypot deployment in the setting of distributed denial-of-service attacks. This paper features one major difference: in \cite{ceker2016deception}, messages are costly, while in  \cite{Carroll2011} they are costless. In other words, defenders in \cite{ceker2016deception} must pay a cost to make normal systems appear to be honeypots or honeypots appear to be normal systems. Because of this cost, separating equilibria emerge in some parameter regimes. Another feature of \cite{ceker2016deception} is that the authors propose a parameter valuation method to obtain the utility functions using existing security evaluations. The authors derive a complete set of closed form solutions.}

Mohammadi et \emph{al}. \citeyear{mohammadi2016game} use signaling games to model the decision processes of a fake avatar that is defending
a social network from attack. Figure \ref{fig:avatar} depicts the
model using the signaling game formulated in Fig. \ref{fig:signaling}. 
A user in the network may be normal or compromised. It interacts with an avatar in suspicious or not suspicious ways, and this interaction is taken to be a signaling-game message. The fake avatar observes this interaction, and forms a belief about the real type of the user. The avatar then decides whether to raise an alert. \cite{mohammadi2016game} is not a cheap-talk game, because raising an alert is assumed to be more costly than not raising an alert. But this extra cost of raising an alert applies regardless of the actual user type, so it does not take the form of a ``lying cost,'' as is the case in \cite{ceker2016deception}. Due to this different structure, the game in \cite{mohammadi2016game} does not support separating equilibria. In the pooling equilibria, the fake avatar alerts the
system of an attack if the prior probability that the user is an attacker
is sufficiently high \cite{mohammadi2016game}.

Pawlick and Zhu present a final perspective on honeypot deployment using signaling games \citeyear{pawlick2018modeling}. They use a cheap-talk game with binary type, message, and action spaces. The unique element, however, is a ``detector'' which gives off probabilistic evidence when the sender misrepresents his type. They call this model a \emph{signaling game with evidence}  \citeyear{pawlick2018modeling,pawlick2015deception}.  This model is capable of nesting cheap-talk signaling games (\emph{c.f.}, \cite{crawford1982strategic}) and games of verifiable disclosure (\emph{c.f}, \cite{milgrom1981good}) as special cases. The relevant equilibrium concept is an extended version of PBNE. The authors
find that the ability to detect honeypots always improves the utility
of the attacker, but surprisingly sometimes also improves the utility
of the defender. This suggests that in certain parameter regions,
it is advantageous for the defender to use imperfect deception. They also find that detectors which prioritize high true-positive rates over low false-positive rates encourage truthful signaling. Finally, they show that receivers prefer equal-error-rate detectors \cite{pawlick2018modeling}.

P\'{i}bil et \emph{al. }\citeyear{pibil2012game} and Kiekintveld et
\emph{al}. \citeyear{kiekintveld2015game} create a different type of imperfect information
model called a honeypot selection game. The honeypot selection game
classifies targets into different levels of importance. These are
assigned corresponding utility values for the attacker, and the defender
plays to minimize the attacker value. The defender disguises the honeypots
by choosing the utility value that they appear to have. A configuration
of the network is represented by a vector of the real values of each
real machine, and the simulated values of the honeypots. The Bayesian Nash equilibrium can be found using a
linear program. The authors examine the defender's optimal strategy
and equilibrium utility as a function of the number of honeypots that
he deploys. Keintveld et \emph{al}. also study an extension of the
game that allows attackers to probe potential targets before choosing
whether to attack. The probe returns accurate or inaccurate responses
with some probability. The game is again zero-sum, and the analysis
proceeds using a linear program \cite{pibil2012game,kiekintveld2015game}.

\subsection{Attacker Engagement}

\label{sub:Attacker-Engagement}
\begin{figure}
\begin{centering}
\includegraphics[width=0.7\columnwidth]{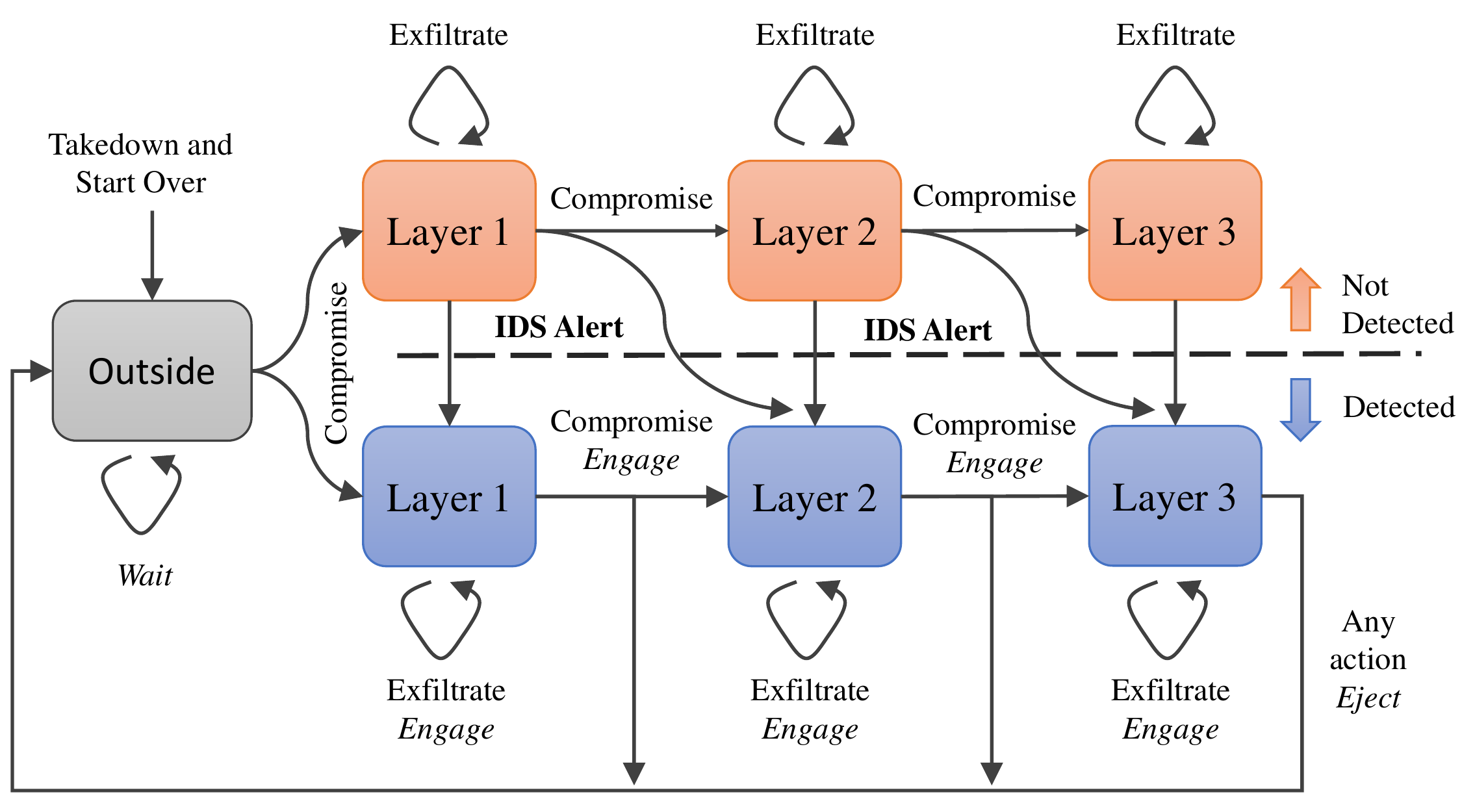}
\par\end{centering}

\caption{\label{fig:engage}Hor\'{a}k et \emph{al}. \citeyear{horak2017manipulating}
model attacker engagement using a one-sided partially observable stochastic
game. From left to right, states correspond to layers of a network.
In the states at the top of the figure, the attacker is undetected.
When the IDS detects the attacker, the state moves to the bottom of
the figure. Defender actions are shown in italics. The defender decides
when to engage the attacker and gather information rather than eject
the attacker.}
\end{figure}
Most game-theoretic models of deception are either static games (such
as simultaneous move Nash games) or single-shot dynamic games (such
as Stackelberg games or signaling games). But some initial work has
investigated multiple-period games. We use ``dynamic'' to refer
to games having multiple periods, and we call these interactions attacker
engagement.

\cite{zhuang2010modeling} create
a multiple-period signaling game that concerns deception and resource
allocation. This paper is about homeland security in general, and
only briefly mentions computer security. Still, it is interesting
due to its technical setup and insights. In each period,
the defender chooses to invest in either short term \textquotedblleft expense\textquotedblright{}
or long term \textquotedblleft capital\textquotedblright{} investments.
In addition, the defender chooses how much of this to reveal to the
attacker using a signal. The defender may signal truthfully or deceptively,
or she may act secretively, \emph{i.e.}, not reveal any information.
The attacker observes the defender\textquoteright s signal, updates
his belief, and chooses whether to attack. Games are constructed in
which the attacker can and cannot observe previous defender investment
strategies. The games are solved through backwards induction. In several
cases, the defender can deceive the attacker, often by mimicking the
strategy of the opposite defender type. For example, when the attacker
cannot observe previous defender strategies and has a high cost of
attack, the defender can deter an attack by choosing an expensive
short term defense rather than long term investment. Zhuang et \emph{al.} also discuss remaining challenges involved in the area of equilibrium
selection \citeyear{zhuang2010modeling}. 

Durkota et \emph{al}. \citeyear{durkota2015optimal} use attack graphs
to represent attacker strategies and develop network-hardening methods
to increase security. They propose a Stackelberg game in which the
defender leads by placing honeypots in the network. The attacker follows
and has knowledge of the number of honeypots (not their identities).
Attack graphs are used to capture the adversary's multiple-round strategies.
Durkota et \emph{al}. argue that vulnerability databases can be used
to contruct a menu of possible attack graphs which the adversary can
employ\textcolor{black}{, assuming that the defender knows which systems, software, etc. are present in the network}. The defense strategy is obtained using Markov decision processes. One contribution
of the paper is an efficient algorithm to solve the Markov decision process using policy
search with pruning \cite{durkota2015optimal}. For the purposes of
the classification in Section \ref{sec:Taxonomy}, it is interesting
to note that the paper builds upon the foundation of honeypot deployment,
but adds the dynamic element of multiple-round attack graphs. 

Hor\'{a}k et \emph{al}. \citeyear{horak2017manipulating} model the penetration of an attacker into a network using a one-sided
partially observable stochastic game. Unlike most approaches to network
security, Hor\'{a}k et \emph{al}. consider the defender rather than the attacker to be the
informed player. Instead of ejecting an attacker
immediately, the defender decides how long to observe the attacker
in order to gather information. Figure \ref{fig:engage} depicts the
set of states and transitions. The attacker compromises the network
in layers and chooses when to exfiltrate data without knowing for
sure whether she has been detected. In equilibrium, the defender's
optimal strategy is to keep the attacker inside the network while
the attacker is confident that he has not yet been detected, and to
eject the attacker from the network otherwise. The authors also evaluate
the robustness of the solution if the attacker does not know the capabilities
of the IDS. While this paper studies a simplified network structure, it offers an uncommon analysis of the
optimal strategies for a powerful and observant defender \cite{horak2017manipulating}.

\section{Taxonomy}

\label{sec:Taxonomy}
The U.S. Department of Defense has
highlighted the need for ``the construction of a common language
and a set of basic concepts about which the security community can
develop a shared understanding'' \cite{mitre2010science}. \textcolor{black}{More specifically, the  term deception has  been  employed
broadly and with a variety of meanings. A ``common language and a shared set of basic concepts'' would allow deception researchers to better collaborate. It would also further the development of cybersecurity and privacy as a science. Finally, a precise taxonomy  would allow game theorists to carefully select models that capture the unique features of each different type of deception.}

In Section \ref{sec:Literature-Survey}, we sorted deception into
six different types: perturbation, moving target defense, obfuscation, mixing, honey-x,
and attacker engagement. In this section, we first give general descriptions
of the six types. Then we construct a taxonomy which precisely defines
each of the types using game-theoretic principles. 
\begin{enumerate}
\item In the area of privacy, deception can limit leakage of sensitive information
though the use of noise. This type of deception is often called \emph{perturbation}
(Subsection \ref{sub:Perturbation}).\label{enu:pert} 
\item Deception can limit the effectiveness of attacker reconnaissance through
techniques such as randomization and reconfiguration of networks,
assets, and defense tools. This is called \emph{moving target defense}
(Subsection \ref{sub:Moving-Target-Defense}).\label{enu:mtd}
\item Deception can waste effort and resources of attackers by directing
them to decoy targets rather than real assets, and can protect privacy
by revealing useless information aside real information. This is called
\emph{obfuscation} (Subsection \ref{sub:Obfuscation}).\label{enu:obf} 
\item Deception can use exchange systems such as mix networks and mix zones
to prevent linkability. We call this type of deception \emph{mixing}
(Subsection \ref{sub:Mixing}).\label{enu:mix} 
\item Deception can draw attackers towards specific systems (such as honeypots)
by disguising these systems as valuable network assets. In order to
include honeynets, honey-users, \emph{etc.}, we call this type of
deception \emph{honey-x} (Subsection \ref{sub:Honey-X}).\label{enu:hon}
\item Deception can use feedback to dynamically influence attackers over
an extended period of time, in order to waste their resources and
gather intelligence about them. We call this \emph{attacker engagement}
(Subsection \ref{sub:Attacker-Engagement}).\label{enu:atkeng}
\end{enumerate}


\subsection{Detailed Definition of Each Type of Deception}

\label{sub:Definition-of-Each}A taxonomy is ``a collection of controlled
vocabulary terms organized into a hierarchical structure'' in which
each term ``is in one or more parent/child (broader/narrower) relationships
to other terms in the taxonomy'' \cite{niso2005vocabularies}. 
Traditionally, each parent is called a \emph{genus}, and each child is called a \emph{species} \cite{chisholm1911taxonomy}.
Each species is demarcated from other members of its genus by a \emph{specific
difference}. 
The most fine-grained species in a taxonomy are sometimes called the \emph{infimae species}. In our taxonomy, we
can now identify the ``types'' of deception more formally as infimae
species.

Since the purpose of our taxonomy is to identify appropriate game-theoretic
models for each infima species of deception,
the specific differences that we use are related to game theory. (See Table \ref{tab:summaryGames}.) These
are the principles of \emph{private information} $\Theta$, \emph{players} $\mathcal{P}$, \emph{actions} $\mathcal{A}$, and \emph{time-horizon} $\mathcal{T}.$

\subsubsection{Private Information}

\label{sub:Incentives}
\begin{figure}
\begin{centering}
\begin{minipage}[t]{1\columnwidth}%
\begin{center}
\includegraphics[width=0.5\columnwidth]{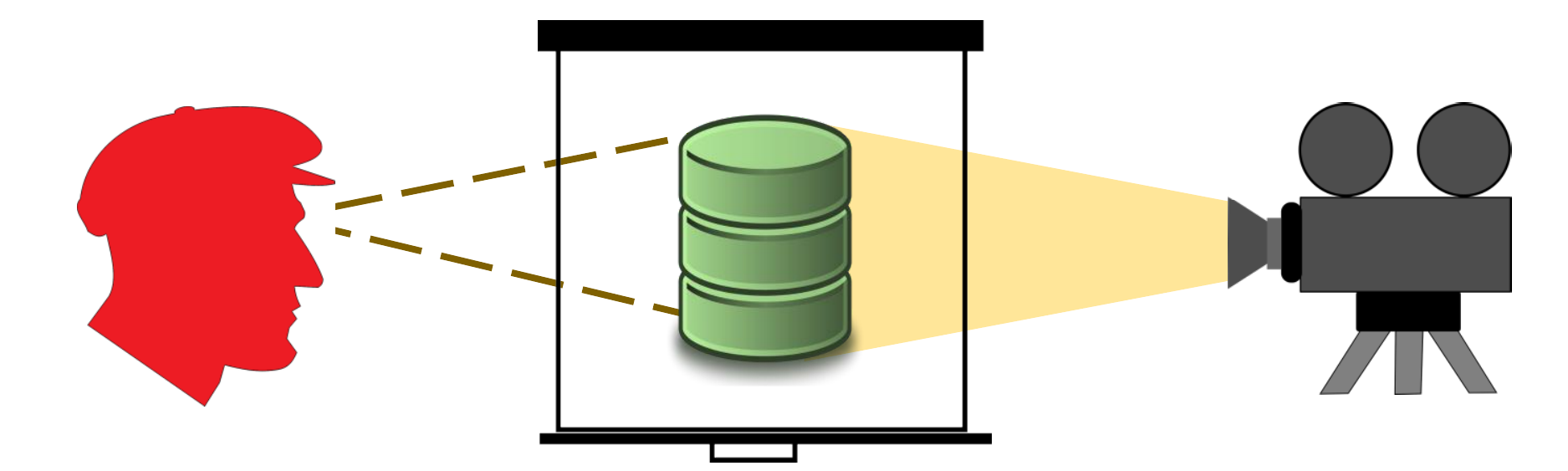}\caption{\label{fig:incentives1}We use the term \emph{mimesis} to signify
the creation of a specific false belief. In this example, a defender
makes an adversary believe that a database exists when, in fact, it
does not exist.}

\par\end{center}%
\end{minipage}
\par\end{centering}

\begin{centering}
\medskip{}

\par\end{centering}

\centering{}%
\begin{minipage}[t]{1\columnwidth}%
\begin{center}
\includegraphics[width=0.5\columnwidth]{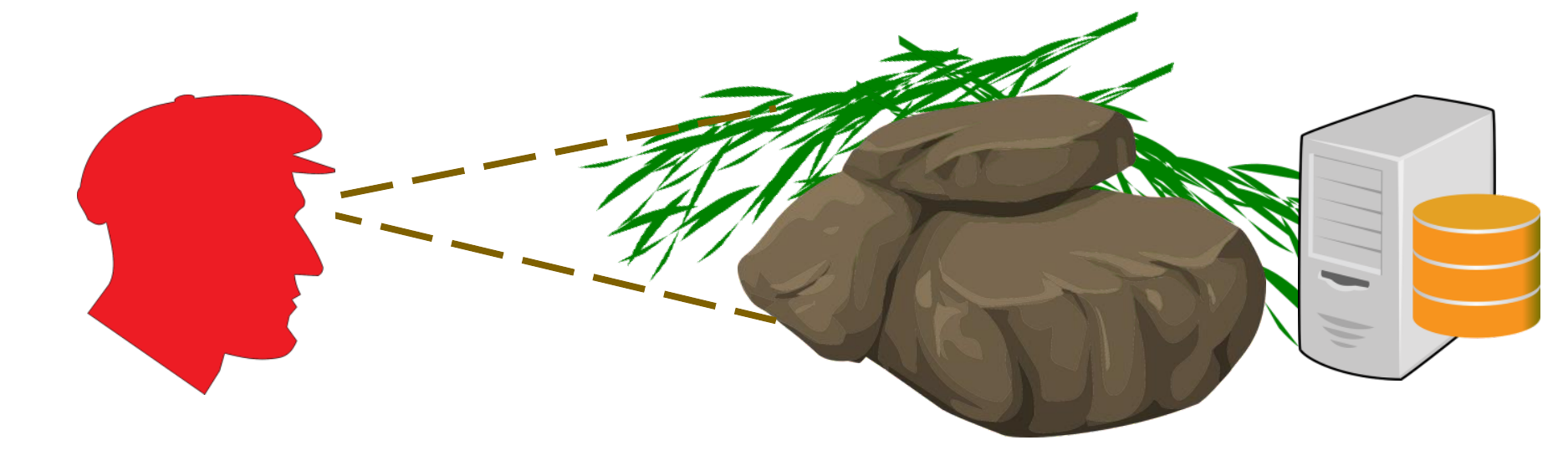}\caption{\label{fig:incentives2}We use the term \emph{crypsis }to refer to
preventing an adversary from acquiring a true belief. In this example,
a defender hides the true existence of a database from an adversary.}

\par\end{center}%
\end{minipage}
\end{figure}

\textcolor{black}{One of these principles is the private information. To deceive is to intentionally cause another agent
either \textquotedblleft to acquire or continue to have a false belief\textquotedblright{}
or \textquotedblleft to be prevented from acquiring or cease to have
a true belief\textquotedblright{} \cite{stanford2016deception}. These
two categories are quite different. Figures \ref{fig:incentives1}-\ref{fig:incentives2}
depict both categories. The first category involves instilling in
another a specific falsity (Fig. \ref{fig:incentives1}). In the language
of game theory, this requires the creation of a belief. 
In signaling games or partially observable stochastic games, agents maintain beliefs
over the private information $\Theta$ of other agents. Deceptive actions manipulate
these beliefs to create traps or decoys.} 
On the other hand, the second
category of deception involves hiding a true belief (Fig. \ref{fig:incentives2}).
This could be employed in cybersecurity to hide a weakness, or it
could be deployed in privacy to protect sensitive information. In
either case, the defender maximizes her utility function by causing
attackers to obtain \textcolor{black}{noisy or uncertain} 
information about the state of reality.

\textcolor{black}{Rothstein and Whaley have distinguised between these two categories as simulation (or ``showing the false'') and dissimulation (or ``hiding the real'') \citeyear{rothstein2013art}. Under simulation, they include subcategories called masking, repackaging, and dazzling. Under dissimulation, they include mimicking, inventing, and decoying. While the distinction between hiding the real and showing the false is appropriate for our purposes, we prefer not to adopt the same subcategories for several reasons. In particular, \cite{rothstein2013art} includes decoying as a type of ``showing the false.'' But we prefer to label it as ``hiding the real.'' In the literature that we review, decoys are simulated entities, but they are used mostly to hide the presence of the real entity\footnote{\textcolor{black}{The use of decoys pertains to obfuscation (Subsection \ref{sub:Obfuscation}). We place technologies such as honeypots (Subsection \ref{sub:Honey-X}) in a separate category, because they are designed to attract attacker attention, not simply to hide the presence of real systems.}}.}

Therefore, we prefer a new terminology. Biologists distinguish hiding the real and showing the false 
 using the terms \emph{crypsis }and \emph{mimesis}.
 ``In the former... an animal resembles some object which is of
no interest to its enemy, and in doing so is concealed; in the latter...
an animal resembles an object which is well known... and in so doing
becomes conspicuous'' \cite{cott1940adaptive}. We adopt these terms
to signify the corresponding categories of cybersecurity deception.

\subsubsection{Actors}

\label{sub:Players}
\begin{figure}
\begin{centering}
\begin{minipage}[t]{1\columnwidth}%
\begin{center}
\includegraphics[width=0.5\columnwidth]{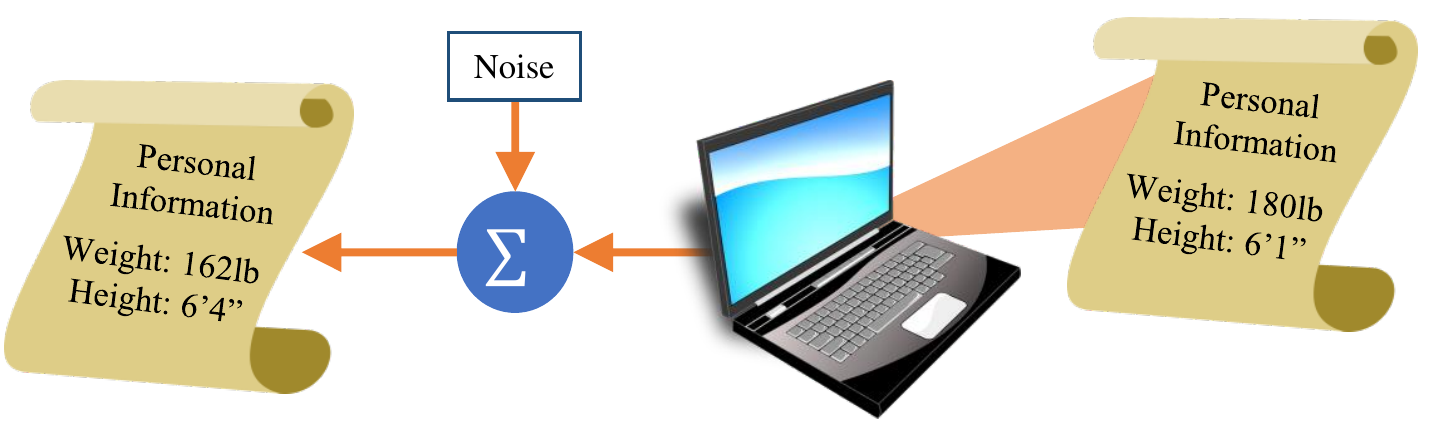}\caption{\label{fig:actors1}Intensive deception. The defender alters the same
object that is being hidden. In this example, the defender adds noise
to private data.}

\par\end{center}%
\end{minipage}
\par\end{centering}

\begin{centering}
\medskip{}

\par\end{centering}

\centering{}%
\begin{minipage}[t]{1\columnwidth}%
\begin{center}
\includegraphics[width=0.5\columnwidth]{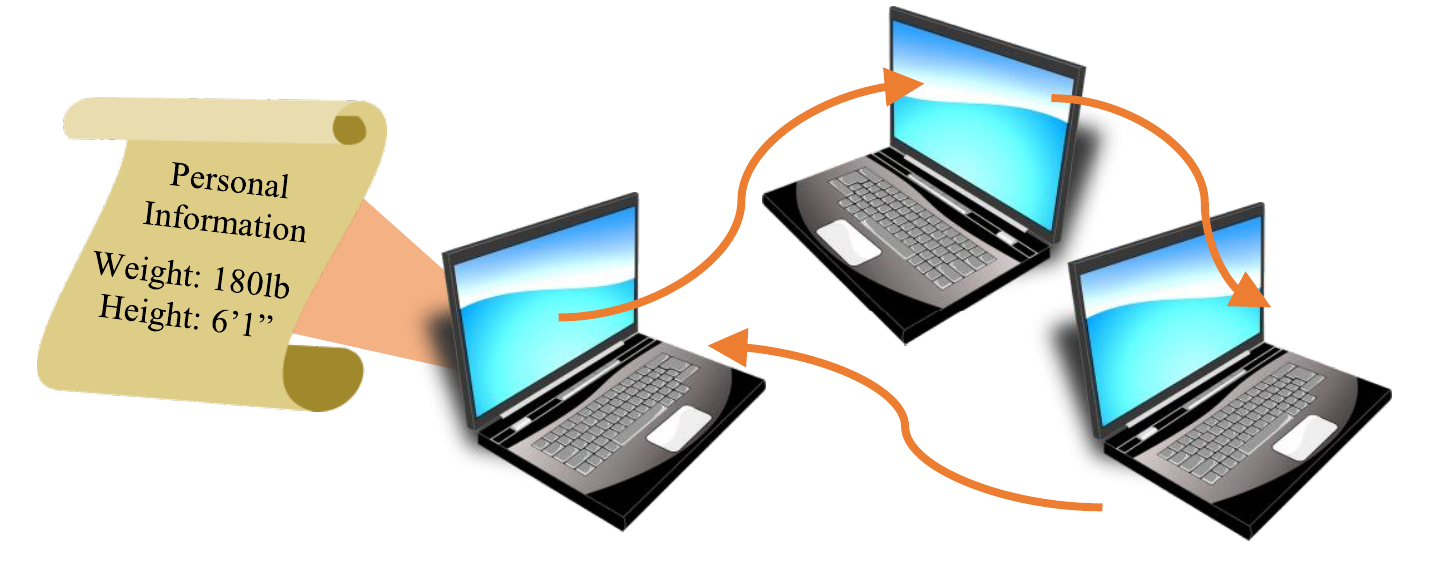}\caption{\label{fig:actors2}Extensive deception. The defender hides an object
using other objects in the environment. In this example, the defender
dynamically changes the location of the private data.}

\par\end{center}%
\end{minipage}
\end{figure}
The next specific difference is the set of actors or players $\mathcal{P}$ involved
in deception.  
\textcolor{black}{Based on the actors involved, cryptic deception can be divided into \emph{intensive} and \emph{extensive}
deception.} (See Fig. \ref{fig:actors1}-\ref{fig:actors2}.)
Intensive deception modifies an
actor (or its own representation) in order to hide it (Fig. \ref{fig:actors1}).
For example, some privacy techniques add noise to data about
a user before publishing it. The user\textquoteright s own data is
modified. By contrast, extensive deception is that category of deception
which hides an actor by using outside actors (Fig. \ref{fig:actors2}).
In cyberspace, examples include mix networks for anonymity. Many messages
enter a network, where they are mixed in a chain of proxy servers
before leaving the network. Any given message is hidden because of
the presence of the other messages.

\subsubsection{Actions}

\label{sub:Actions}Deceivers can act in a number of ways, or actions
$\mathcal{A}.$ 
Within cryptic deception,
we distinguish between deception that uses information and deception
that uses motion. Deception that uses information tends to manipulate
the data released about agents\textquoteright{} properties,
while deception that uses motion either modifies these properties
over time or realizes these properties from a random variable. In
other words, the first category is associated with creating noise,
while the second category is associated with concepts such as agility
and randomization.

\subsubsection{Duration}

\label{sub:Time-Horizon}Within mimetic deception, we distinguish
between static and dynamic scenarios. Dynamic games feature multiple interactions, while static 
games consist of only one interaction\footnote{Here, we consider one-shot games to be static even if they are not
simultaneous-move games.}. Currently, the majority of game-theoretic defensive deceptions use
static models.
The most popular set of mimetic deceptions that are static are honeypots,
honeynets, and honeytokens, to which we refer generically using the
term \emph{honey-x}. We use the name \emph{attacker engagement} to
denote all forms of mimetic deception which are dynamic.

\subsection{Synthesizing the Taxonomy}

\begin{figure*}
\centering{}\includegraphics[width=0.8\textwidth]{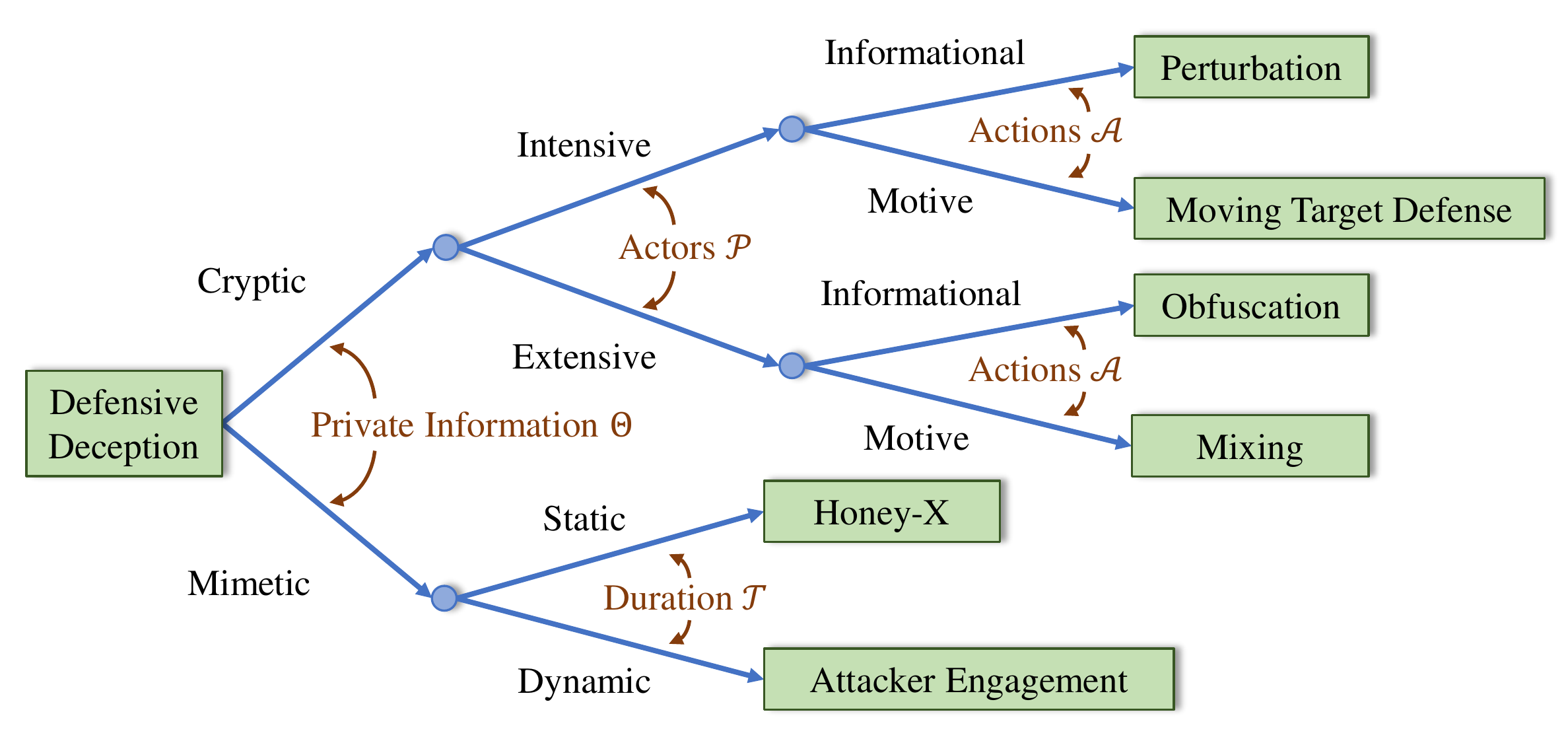}\caption{\label{fig:taxonomy}Tree diagram breakdown of deception into various
species. The specific differences correspond to the game-theoretic
notions of private information, actors, actions, and duration. We describe
a set of infimae species called perturbation, moving target defense,
obfuscation, mixing, honey-x, and attacker engagement. }
\end{figure*}
Figure \ref{fig:taxonomy} uses the specific differences of \textcolor{black}{private information}, 
actors, actions, and duration to create a taxonomy that breaks the
genus of deception into multiple levels of species, ending with the
infimae species of perturbation, moving target defense, obfuscation, \emph{etc}. Based
on the taxonomy, Table \ref{tab:Definitions} lists the definitions
of each of the infimae species. The definitions are unambiguous as
long as the specific differences are clearly defined. They are also mutually exclusive, because we have used mutually exclusive specific differences. Finally, note that the order of application of the specific differences does not matter. This implies that the taxonomy could be also represented
using a binary lattice in four dimensions, one dimension for each
of the specific differences.

\begin{table}%
\centering{}%
\tbl{Infimae Species and Definition\label{tab:Definitions}}{%
{\small{}}%
\begin{tabular}{|c|c|}
\hline 
{\small{}Infima Species} & {\small{}Definition}\tabularnewline
\hline 
\hline 
{\small{}Perturbation} & {\small{}Cryptic, intensive, informational deception}\tabularnewline
\hline 
{\small{}Moving Target Defense} & \multirow{1}{*}{{\small{}Cryptic, intensive, motive deception}}\tabularnewline
\hline 
{\small{}Obfuscation} & {\small{}Cryptic, extensive, informational deception}\tabularnewline
\hline 
{\small{}Mixing} & {\small{}Cryptic, extensive, motive deception}\tabularnewline
\hline 
{\small{}Honey-X} & {\small{}Mimetic, static deception}\tabularnewline
\hline 
{\small{}Attacker Engagement} & \multirow{1}{*}{{\small{}Mimetic, dynamic deception}}\tabularnewline
\hline 
\end{tabular}{\small \par}}
\end{table}%




\subsubsection{Classification of Literature}

In the appendix, Table \ref{tab:justification1} and Table \ref{tab:justification2}
justify the classification of each paper into its infimae species
based on the value of its specific differences. Most classifications
are straightforward. In order to evaluate the structure of the taxonomy, however, consider several border cases.

First, \cite{shokri2015privacy} uses the term ``obfuscation,''
but we have classified the deception in this paper as perturbation.  Noise
is applied to the valuable information itself, so our taxonomy categorizes
it as intrinsic rather than extrinsic. Nevertheless, obfuscation and
perturbation are closely related. If we view each specific difference
of an infimae species as a bit, then we can say that obfuscation and
perturbation have a Hamming distance of one. 

Second, consider \cite{zhu2012deceptive}.
The main idea of the paper is to protect valuable network traffic
by sending external decoy traffic. Therefore, the infimae species
is obfuscation. But the paper also considers mixed strategies, which
implement the agility or motion that is characteristic of moving target defense. Again,
obfuscation and moving target defense have a Hamming distance of one. 

A third border
case is \cite{zhang2010gpath}, which studies
protection of traffic in the \emph{Tor} network. This application
is similar to the deceptive routing problems in \cite{clark2012deceptive,zhu2012deceptive},
which we have classified as obfuscation. Still, we have classified
\cite{zhang2010gpath} as mixing, because \emph{Tor} requires the
participation of other users in the mix network; therefore, it is
extrinsic. 


\subsubsection{Observed Relationships between Models and Infimae Species}

\begin{figure}
\begin{centering}
\begin{minipage}[t]{1\columnwidth}%
\begin{center}
\includegraphics[width=0.7\columnwidth]{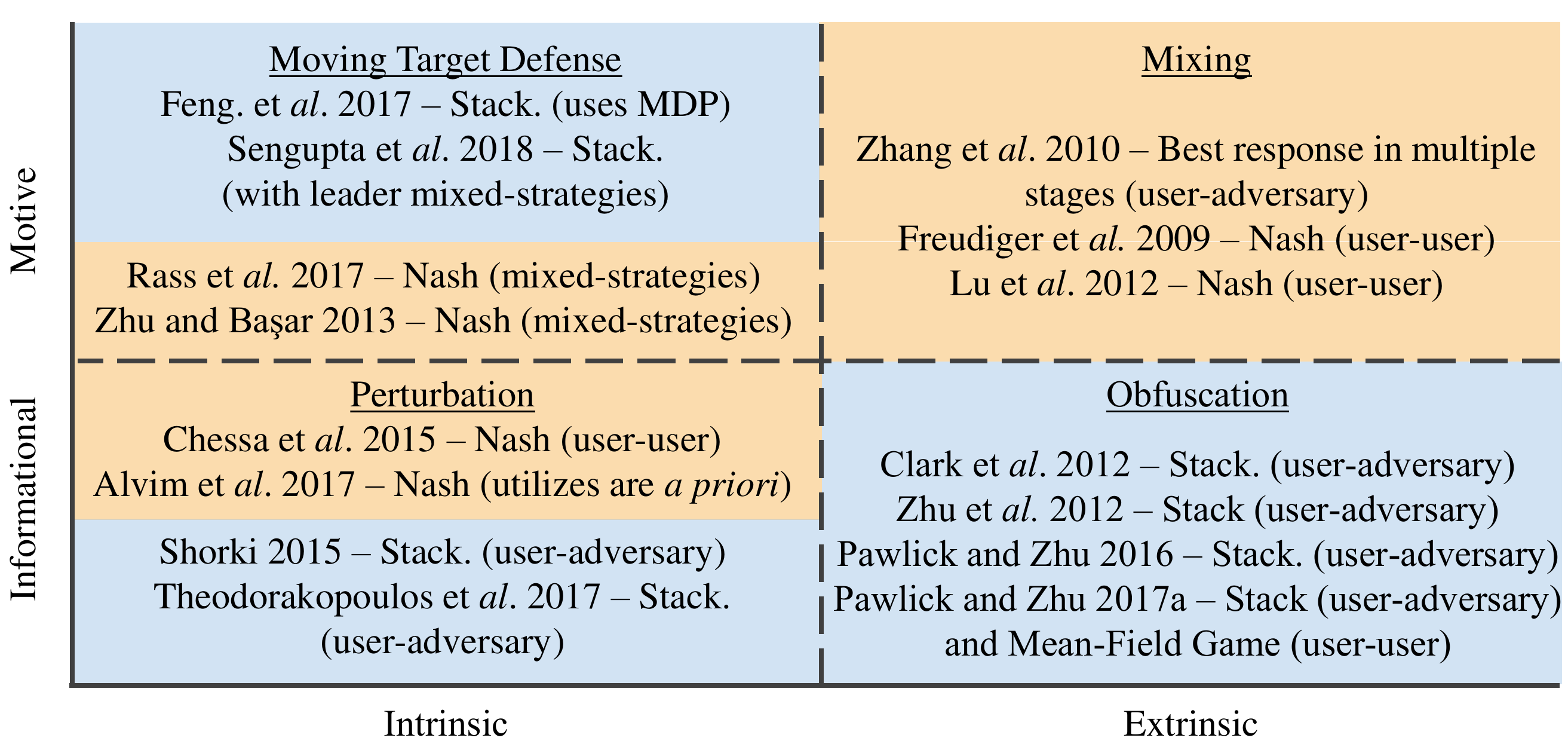}
\par\end{center}

\caption{\label{fig:speciesVmodelsCrypsis1}The game-theoretic model is listed
for each paper in cryptic deception, in order to compare and contrast
the models used for each infimae species. Parenthetical remarks give reasons for the chosen models.}
\end{minipage}
\par\end{centering}

\medskip{}

\centering{}%
\begin{minipage}[t]{1\columnwidth}%
\begin{center}
\includegraphics[width=0.7\columnwidth]{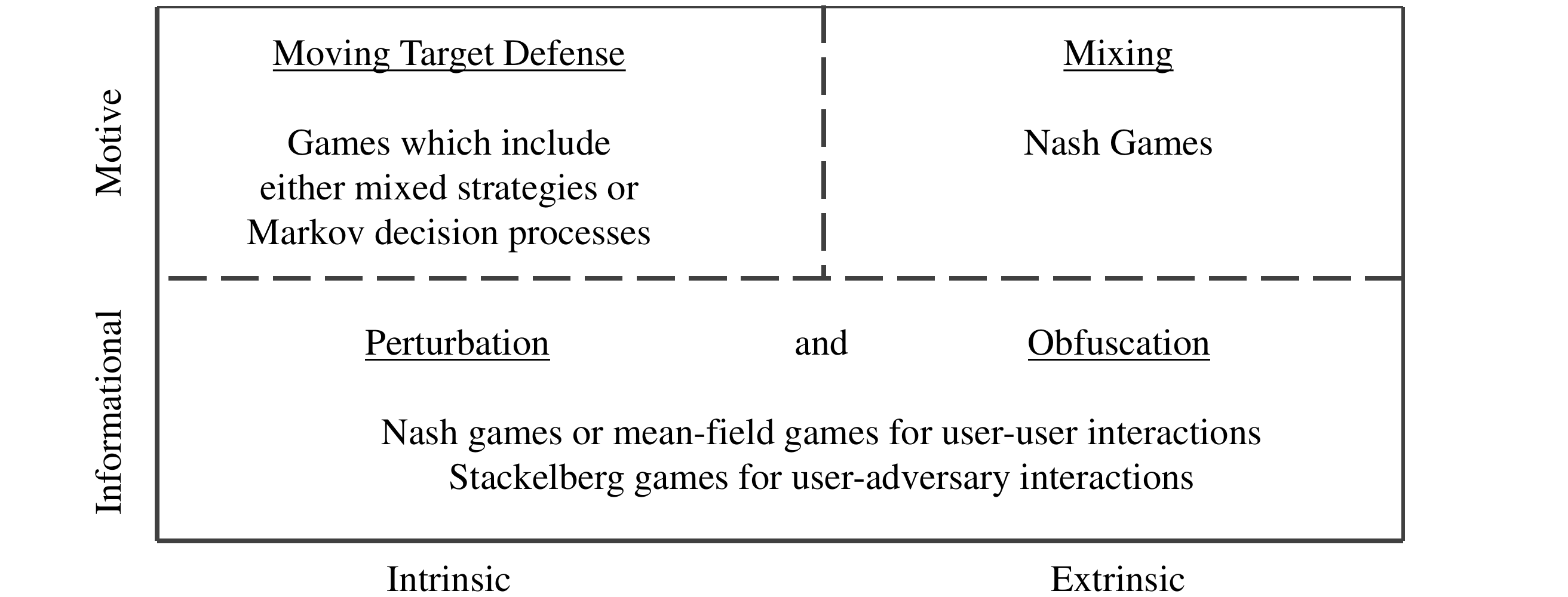}
\par\end{center}

\caption{\label{fig:speciesVmodelsCrypsis2}Several trends are apparent in
the mapping between infimae species and game-theoretic models.}
\end{minipage}
\end{figure}
\textcolor{black}{Since our taxonomy differentiates species of deception based on game theory, it is interesting to see which games were used to model each species. } 
Figure \ref{fig:speciesVmodelsCrypsis1} lists the papers from Section
\ref{sec:Literature-Survey} that study crypsis. 
Most papers use Nash games and Stackelberg games. The papers that use Nash games
(or related concepts) are listed on an orange background, and the
papers that use Stackelberg games are listed on a blue background. 
\textcolor{black}{The figure reveals that authors do not concur on a one-to-one mapping between species and games. This is to be expected, since 1) different games capture different aspects of the same interaction, 2) modeling approaches are still evolving in this nascent field, and 3) the cybersecurity landscape itself is evolving.}

\textcolor{black}{Nevertheless, some trends are evident. These are depicted in Fig. \ref{fig:speciesVmodelsCrypsis2}.} 
First, all
of the papers on mixing use Nash games or related concepts. (Zhang
et \emph{al}. 2010 uses several rounds of best-response which is based
on the idea of Nash equilibrium.) This is logical, because each user
in a mix network or a mix zone simultaneously decides whether to participate.
A second trend spans both perturbation and obfuscation. Papers which
study user-user interactions are modeled using Nash games (or in one
case, a mean-field game), and papers which study user-adversary interactions
are generally modeled using Stackelberg games. Finally, moving target defense tends 
to be modeled in two different ways. One way is to use mixed strategies to model randomness in defensive configurations. 
The second way is to use Markov decision processes to explicitly model
the temporal component of changing defensive characterizations.
\begin{figure}
\begin{centering}
\begin{minipage}[t]{1\columnwidth}%
\begin{center}
\includegraphics[width=0.7\columnwidth]{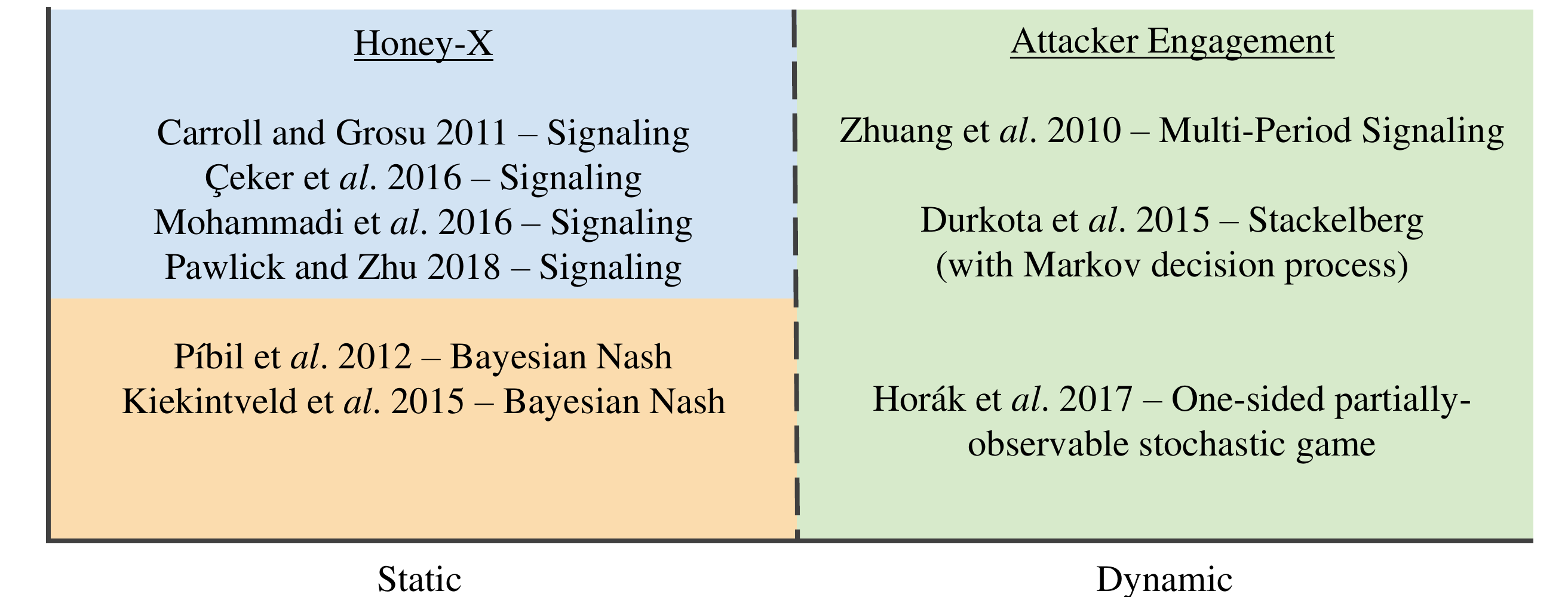}
\par\end{center}

\caption{\label{fig:speciesVmodelsMimesis1}The game-theoretic model is listed
for each paper in mimetic deception, in order to compare and contrast
the models used for each infimae species.}
\end{minipage}
\par\end{centering}

\medskip{}

\begin{minipage}[t]{1\columnwidth}%
\begin{center}
\includegraphics[width=0.7\columnwidth]{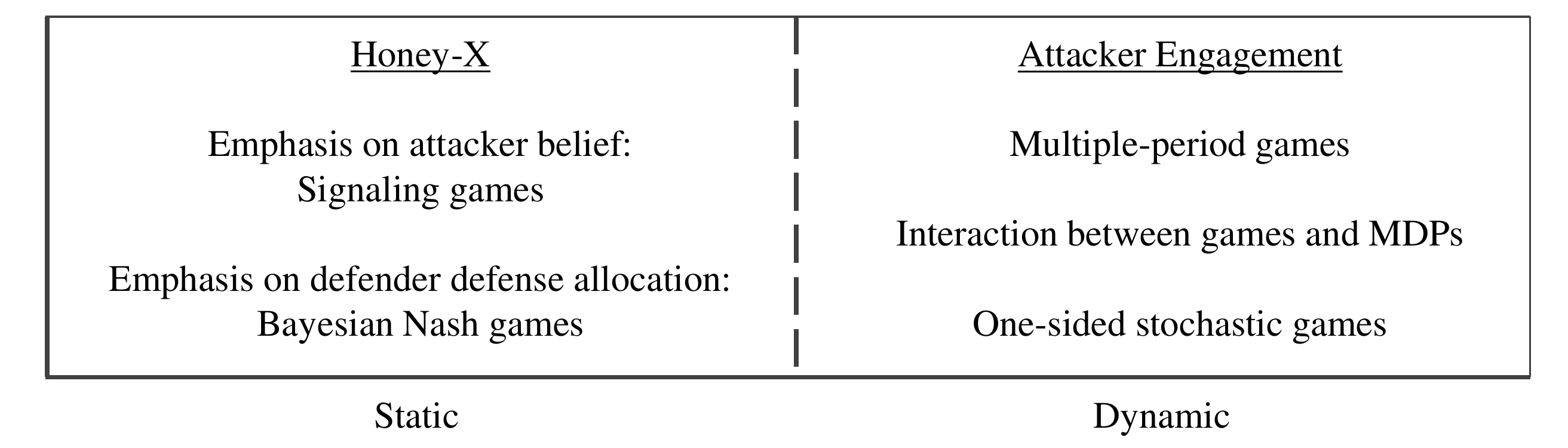}
\par\end{center}

\caption{\label{fig:speciesVmodelsMimesis2}Trends are indicated in each species
of mimesis.}
\end{minipage}
\end{figure}

Figures \ref{fig:speciesVmodelsMimesis1}-\ref{fig:speciesVmodelsMimesis2}
list the papers that study mimesis. 
Papers on the left-hand side study
honey-x (static mimesis), and papers on the right-hand side study
attacker engagement (dynamic mimesis). 
\textcolor{black}{The mapping between deception species and games is not one-to-one.} 
Honey-x is modeled using two
different approaches. One approach uses signaling games, in order
to emphasize the formation of the attacker's belief about whether
systems are normal systems or honeypots. The other approach uses Bayesian
Nash games. This approach follows along the lines of resource allocation
problems, and obtains an overall network configuration that is optimal
for the defender. 

The right-hand sides of Fig. \ref{fig:speciesVmodelsMimesis1}-\ref{fig:speciesVmodelsMimesis2}
list three approaches to attacker engagement. 
\cite{zhuang2010modeling} use a multiple-period game which has a
state that reflects information from past periods. The solution is
obtained using dynamic programming. \cite{durkota2015optimal}
represent a dynamic network attack using a Markov decision process. The attacker deploys
the whole Markov decision process as a Stackelberg follower, and the defender places honeypots
as a Stackelberg leader. Finally, \cite{horak2017manipulating} 
use a one-sided partially-observable Markov decision process. This model is perhaps the
closest of the three to a general \emph{competitive Markov decision process}, in which
a defender and an attacker choose dynamic policies to optimize long-term
utility functions \cite{filar2012competitive}. 

\subsubsection{\textcolor{black}{Preferred Relationships between Models and Infimae Species}}

\begin{figure}

\begin{minipage}[t]{1\columnwidth}%
\begin{center}
\includegraphics[width=0.9\columnwidth]{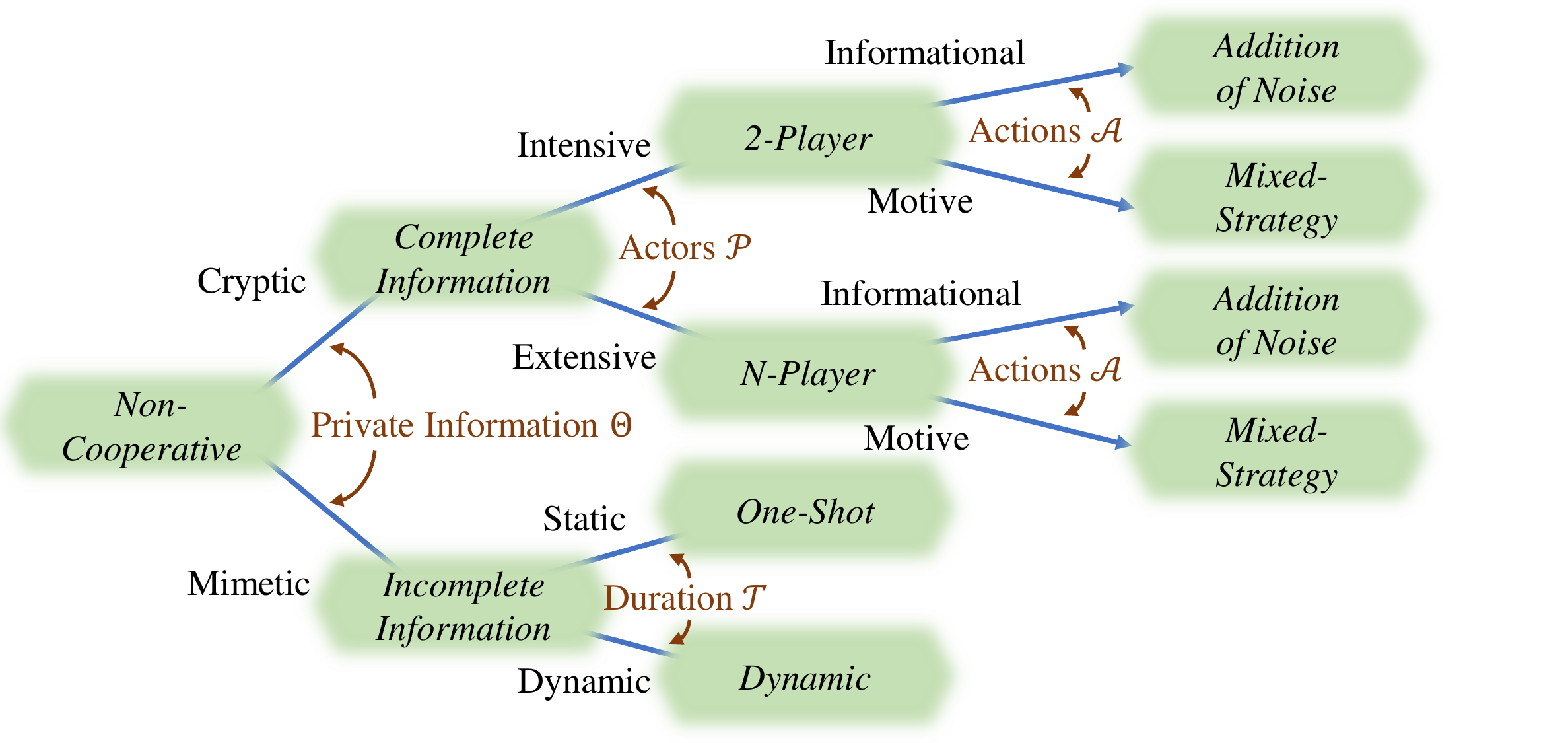}
\par\end{center}

\caption{\label{fig:revisedTaxWithGames}\textcolor{black}{Taxonomy with promising modeling approaches overlaid in a hierarchical fashion. For instance, cryptic, intensive, and motive deception (\emph{i.e.,} moving target defense) can be modeled by non-cooperative, complete-information, two-player games with mixed strategies. Each of the modeling approaches is shown in a blurred shape to indicate that alternate approaches are possible.}}
\end{minipage}

\end{figure}

\textcolor{black}{Within these trends, certain modeling approaches best capture the essential elements of each species of deception. Figure \ref{fig:revisedTaxWithGames} depicts these approaches. 
The root node is non-cooperative games\footnote{Future research can consider cooperative games, but most current studies use non-cooperative models.}. Within non-cooperative games, games with incomplete information are well suited for mimetic deception, since they can explicitly model the false attacker beliefs that defenders attempt to inculcate. One-shot models such as signaling games model honey-x, while dynamic models such as partially-observable stochastic games are necessary to effectively study attacker engagement.}

\textcolor{black}{Within cryptic deception, two players are often sufficient to study intensive deception, in which a single defender attempts to directly deceive an attacker. By contrast, extensive deception involves multiple defenders who compete against each other in addition to competing against an attacker. In perturbation and obfuscation, actions often consist of choosing the variance of noise to add to published data. In mixing and moving target defense, actions often consist of choosing mixed strategies. Through these mixed strategies, defenders probabilistically choose routing paths or network configurations.}

\section{Discussion and Future Directions}

\label{sec:Future-Directions}Our taxonomy and literature survey suggest
a number of open challenges and promising future directions.

\subsection{Mimesis}

In the papers that we survey, cryptic deception predominates over mimetic deception. One
explanation is that crypsis is the goal of all privacy research.  Another explanation is that the straight-forward technique of randomization falls within 
the category of crypsis. Finally, research in mimesis may 
face greater ethical concerns than in crypsis. For instance, ought 
a Chief Information Officer to publish a document which falsely states
the number of data servers that the company runs in order mislead
an adversary? Clearly, there are some aspects of deception that could
infringe on other values such as trustworthiness. These caveats notwithstanding,  
more opportunity remains in mimetic deception.

\subsection{Theoretical advances}

The vast majority of the papers that we have surveyed use Stackelberg
games and Nash games. Many combine these with rich non-strategic models.
For example, \cite{feng2017mtd} combines a Stackelberg game with
a Markov decision process, and \cite{zhu2013game} combines a Stackelberg game with dynamic
systems analysis. Few papers, though, include advanced game-theoretic
models. For instance, the literature that we surveyed did not include
any cooperative games. In addition, dynamic games were not often considered,
partly because general stochastic games are fundamentally difficult
to analyze. This may also reflect the relative newness of research
in defensive deception for cybersecurity and privacy---and the application
of game theory to this area.

\subsection{Practical implementations}

Game theory has been successfully deployed for physical security in several applications. For instance, \cite{pita2008deployed} describes the ARMOR
system, which is a game-theoretic protocol deployed at Los Angeles
International Airport. This protocol uses Stackelberg games. Similarly,
\cite{jain2010software} uses Stackelberg games in order to optimize
the assignment of Federal Air Marshals to U.S. commercial flights.
Yet it is difficult to identify successful implementations of game-theoretic
concepts to cybersecurity. Of course, it is possible that commercial
endeavors use game-theoretic defenses, but prefer not to publish the
results. Still, several challenges frustrate the deployment of game
theory in cybersecurity. 

One obstacle is that \textcolor{black}{some} cybersecurity professionals \textcolor{black}{may be} wary of
so-called \emph{security through obscurity}, \emph{i.e.}, deceptive
security mechanisms that rely only on an attacker's lack of information.
It is true that deceptive mechanisms should be combined with traditional
approaches such as cryptography and access control. In our opinion,
though, game theory provides precisely the right set of tools to offer
security professionals \emph{provable guarantees} on what can be achieved
by deception. Many game theory models assume that the adversary has
full access to the strategy of the defender. Hence, they obtain worst-case
guarantees, and are not undermined if an attacker learns the defender's
strategy. 
\textcolor{black}{Indeed, some government organizations have enthusiastically adopted deception for cybersecurity. For example, researchers at the American National Security Agency (NSA) recently conducted a series of observed red-team exercises involving deception. Data collectors monitored the actions and rationale of the attackers and defenders \cite{fergusonwalter2017friend}. Such exercises are invaluable in making sure that game-theoretic models accurately capture elements of human psychology.}

A second obstacle is, paradoxically, the high demand for technical
network security professionals. Since industry and government organizations
have a high demand for security analysts, and since these analysts
need to constantly monitor network traffic, it is challenging to establish
collaborations between the analysts and academic researchers with
backgrounds in game theory. This collaboration is crucial in order
to ensure that game theorists solve relevant problems. On the other
hand, work in game theory also offers the possibility to relieve the
burden on analysts. For example, game theory can be used to optimally
design IDS alerts in order to limit the number
of false positives that must be analyzed.

\subsection{Interdisciplinary Security}

Our paper has analyzed deception as a quantitative science. Deception,
however, is interdisciplinary. Economics research 
offers well-developed game-theoretic models. \textcolor{black}{Often, however, the challenge is to apply these models in accurate, domain-specific ways.} Work in experimental
economics emphasizes behavioral or sub-rational aspects of deception. 
These aspects are critical for applications in which the attacker does not play 
the theoretically-optimal strategy. Additionally, psychology can be employed 
to analyze attacker preferences and develop accurate threat models. \textcolor{black}{Finally, criminology can be useful to detect signs of deception by humans. As attacks become increasingly automated, though, the effectiveness of both psychology and criminology will have to be reevaluated.}

\subsection{Conclusion}

The next generation of cybersecurity and privacy techniques will leverage
tools commonly employed by attackers for the purpose of defense. For
applications ranging from protection of civil liberties to network
security to defense of the Internet of Battle Things, defensive deception
will play a principal role. We have developed a taxonomy of defensive
deception for cybersecurity and privacy viewed through the lens of
game theory. This taxonomy provides a scientific foundation for future
defensive deception research and a common language that can be used
to conceptualize defensive deception. This work also provides
a menu of game-theoretic models and defensive deception techniques
that can be leveraged for future research. Finally, we have summarized
many of the contributions of game theory to the various species of
deception over the past ten years. We hope that these will provide
a conceptual basis for the next stage of research in defensive deception
in cybersecurity and privacy.

\appendix
\section*{APPENDIX}
\setcounter{section}{1}

Table \ref{tab:justification1} and Table \ref{tab:justification2} list our 
classification justifications for papers in cryptic and mimetic deception, respectively.

\begin{table}%
\tbl{\label{tab:justification1}Classification Justification for Papers
in Cryptic Deception}{%
\centering{}{\small{}}%
\begin{tabular}{|c|c|l|}
\hline 
{\small{}Authors and Year} & {\small{}Category} & {\small{}Justification}\tabularnewline
\hline 
\hline 
{\small{}Chessa et }\emph{\small{}al}{\small{}. 2015 } & {\small{}Cryptic} & {\small{}The objective is to hide private data. }\tabularnewline
\cline{2-3} 
 & {\small{}Intrinsic} & {\small{}The private data itself is modified.}\tabularnewline
\cline{2-3} 
 & {\small{}Informational} & {\small{}Noise is added to a quantity.}\tabularnewline
\hline 
{\small{}Shokri 2015 } & {\small{}Cryptic} & {\small{}The objective is to hide private data. }\tabularnewline
\cline{2-3} 
 & {\small{}Intrinsic} & {\small{}The private data itself is modified.}\tabularnewline
\cline{2-3} 
 & {\small{}Informational} & {\small{}Noise is added to a quantity.}\tabularnewline
\hline 
{\small{}Alvim et }\emph{\small{}al}{\small{}. 2017 } & {\small{}Cryptic} & {\small{}The objective is to hide private data. }\tabularnewline
\cline{2-3} 
 & {\small{}Intrinsic} & {\small{}The private data itself is modified.}\tabularnewline
\cline{2-3} 
 & {\small{}Informational} & {\small{}Noise is added to a quantity.}\tabularnewline
\hline 
{\small{}Theodorakopoulos } & {\small{}Cryptic} & {\small{}The objective is to hide location.}\tabularnewline
\cline{2-3} 
{\small{}et }\emph{\small{}al}{\small{}. 2014 } & {\small{}Intrinsic} & {\small{}The location itself is modified.}\tabularnewline
\cline{2-3} 
 & {\small{}Informational} & {\small{}Noise is added to the location.}\tabularnewline
\hline 
{\small{}Rass et }\emph{\small{}al}{\small{}. 2017 } & {\small{}Cryptic} & {\small{}The goal is to hide configuration.}\tabularnewline
\cline{2-3} 
 & {\small{}Intrinsic} & {\small{}The configuration is centrally-controlled.}\tabularnewline
\cline{2-3} 
 & {\small{}Motive} & {\small{}The configuration is varied over time.}\tabularnewline
\hline 
\textcolor{black}{{\small{}Sengupta et }\emph{\small{}al}{\small{}. 2018 }} & \textcolor{black}{{\small{}Cryptic}} & \textcolor{black}{{\small{}The goal is to hide IDS locations.}}\tabularnewline
\cline{2-3} 
 & \textcolor{black}{{\small{}Intrinsic}} & \textcolor{black}{{\small{}The locations are chosen centrally.}}\tabularnewline
\cline{2-3} 
 & \textcolor{black}{{\small{}Motive}} & \textcolor{black}{{\small{}The locations are is realized randomly.}}\tabularnewline
\hline 
{\small{}Zhu \& Ba\c{s}ar 2013 } & {\small{}Cryptic} & {\small{}The goal is to hide vulnerability locations.}\tabularnewline
\cline{2-3} 
 & {\small{}Intrinsic} & {\small{}The configuration is centrally-controlled.}\tabularnewline
\cline{2-3} 
 & {\small{}Motive} & {\small{}The configuration is varied over time.}\tabularnewline
\hline 
{\small{}Feng et }\emph{\small{}al}{\small{}. 2017} & {\small{}Cryptic} & {\small{}The goal is to hide a resource configuration.}\tabularnewline
\cline{2-3} 
 & {\small{}Intrinsic} & {\small{}The configuration is centrally-controlled.}\tabularnewline
\cline{2-3} 
 & {\small{}Motive} & {\small{}The configuration is varied over time.}\tabularnewline
\hline 
{\small{}Clark et }\emph{\small{}al}{\small{}. 2012 } & {\small{}Cryptic} & {\small{}The goal is to hide the network traffic path.}\tabularnewline
\cline{2-3} 
 & {\small{}Extrinsic} & {\small{}Traffic is hidden using other useless traffic.}\tabularnewline
\cline{2-3} 
 & {\small{}Informational} & {\small{}This uses uncertainty rather than motion.}\tabularnewline
\hline 
{\small{}Zhu et }\emph{\small{}al}{\small{}. 2012 } & {\small{}Cryptic} & {\small{}The goal is to hide the network traffic path.}\tabularnewline
\cline{2-3} 
 & {\small{}Extrinsic} & {\small{}Traffic is hidden using other useless traffic.}\tabularnewline
\cline{2-3} 
 & {\small{}Informational} & {\small{}This uses uncertainty rather than motion.}\tabularnewline
\hline 
{\small{}Pawlick \& } & {\small{}Cryptic} & {\small{}The goal is to hide true user behavior.}\tabularnewline
\cline{2-3} 
\multirow{1}{*}{{\small{}Zhu 2016 }} & {\small{}Extrinsic} & {\small{}True behavior is hidden in distracting actions.}\tabularnewline
\cline{2-3} 
 & {\small{}Informational} & {\small{}This uses uncertainty rather than motion.}\tabularnewline
\hline 
{\small{}Pawlick \& } & {\small{}Cryptic} & {\small{}The goal is to hide true user behavior.}\tabularnewline
\cline{2-3} 
{\small{}Zhu 2017a } & {\small{}Extrinsic} & {\small{}True behavior is hidden in distracting actions.}\tabularnewline
\cline{2-3} 
 & {\small{}Informational} & {\small{}This uses uncertainty rather than motion.}\tabularnewline
\hline 
{\small{}Zhang et }\emph{\small{}al}{\small{}. 2010 } & {\small{}Cryptic} & {\small{}The goal is to hide entry-exit node pairings.}\tabularnewline
\cline{2-3} 
 & {\small{}Extrinsic} & {\small{}The mix network relies on other nodes.}\tabularnewline
\cline{2-3} 
 & {\small{}Motive} & {\small{}Exit and entry nodes are realized randomly.}\tabularnewline
\hline 
{\small{}Freudiger } & {\small{}Cryptic} & {\small{}The goal is to hide user-pseudonym links.}\tabularnewline
\cline{2-3} 
{\small{}et }\emph{\small{}al}{\small{}. 2009 } & {\small{}Extrinsic} & {\small{}Swaps require multiple participants.}\tabularnewline
\cline{2-3} 
 & {\small{}Motive} & {\small{}Hiding links depends on multiple exchanges.}\tabularnewline
\hline 
{\small{}Lu et }\emph{\small{}al. }{\small{}2012 } & {\small{}Cryptic} & {\small{}The goal is to hide user-pseudonym links}\tabularnewline
\cline{2-3} 
 & {\small{}Extrinsic} & {\small{}Swaps require multiple participants.}\tabularnewline
\cline{2-3} 
 & {\small{}Motive} & {\small{}Hiding links depends on multiple exchanges.}\tabularnewline
\hline 
\end{tabular}{\small \par}}
\end{table}%

\begin{table}%
\tbl{\label{tab:justification2}Classification Justification for Papers in Mimetic Deception}{%
\centering{}{\small{}}%
\begin{tabular}{|c|c|l|}
\hline 
{\small{}Authors and Year} & {\small{}Category} & {\small{}Justification}\tabularnewline
\hline 
\hline 
{\small{}Carroll \&} & {\small{}Mimetic}  & {\small{}Honeypots appear to be normal systems.}\tabularnewline
\cline{2-3} 
{\small{}Grosu 2011} & {\small{}Static}  & {\small{}The interaction is a one-shot game.}\tabularnewline
\hline 
\textcolor{black}{\small{}\c{C}eker et \emph{al}. 2016} & \textcolor{black}{\small{}Mimetic} & \textcolor{black}{{\small{}Honeypots appear to be normal systems.}}\tabularnewline
\cline{2-3} 
 & \textcolor{black}{\small{}Static} & {\small{}\textcolor{black}{\small{}The interaction is a one-shot game.}}\tabularnewline
\hline 
{\small{}Mohammadi} & {\small{}Mimetic}  & {\small{}Honeypots appear to be normal systems.}\tabularnewline
\cline{2-3} 
et \emph{al\@.}{\small{} 2016} & {\small{}Static}  & {\small{}The interaction is a one-shot game.}\tabularnewline
\hline 
{\small{}Pawlick \&} & {\small{}Mimetic}  & {\small{}Honeypots appear to be normal systems.}\tabularnewline
\cline{2-3} 
{\small{}Zhu 2018} & {\small{}Static}  & {\small{}The interaction is a one-shot game.}\tabularnewline
\hline 
{\small{}P\'{i}bil et }\emph{\small{}al}{\small{}. 2012} & {\small{}Mimetic}  & {\small{}Honeypots appear to be normal systems.}\tabularnewline
\cline{2-3} 
 & {\small{}Static}  & {\small{}The interaction is a one-shot game.}\tabularnewline
\hline 
{\small{}Kiekintveld} & {\small{}Mimetic}  & {\small{}Honeypots appear to be normal systems.}\tabularnewline
\cline{2-3} 
{\small{} et }\emph{\small{}al}{\small{}. 2015} & {\small{}Static}  & {\small{}The interaction is a one-shot game.}\tabularnewline
\hline 
{\small{}Zhuang} & {\small{}Mimetic}  & {\small{}Defender types can signal the opposite type.}\tabularnewline
\cline{2-3} 
{\small{}et }\emph{\small{}al}{\small{}. 2010} & {\small{}Dynamic}  & {\small{}The game occurs over multiple periods.}\tabularnewline
\hline 
{\small{}Durkota} & {\small{}Mimetic}  & {\small{}Honeypots appear to be normal systems.}\tabularnewline
\cline{2-3} 
{\small{}et }\emph{\small{}al}{\small{}. 2015} & {\small{}Dynamic}  & {\small{}Attacks are multi-stage.}\tabularnewline
\hline 
{\small{}Hor\'{a}k} & {\small{}Mimetic}  & {\small{}The attacker believes he is undetected.}\tabularnewline
\cline{2-3} 
{\small{}et }\emph{\small{}al}{\small{}. 2017} & {\small{}Dynamic}  & {\small{}The game is infinite-horizon.}\tabularnewline
\hline 
\end{tabular}}
\end{table}

\clearpage 
\bibliographystyle{ACM-Reference-Format-Journals}
\bibliography{PDoSjBib}


\begin{thebibliography}{00}


\ifx \showCODEN    \undefined \def \showCODEN     #1{\unskip}     \fi
\ifx \showDOI      \undefined \def \showDOI       #1{{\tt DOI:}\penalty0{#1}\ }
  \fi
\ifx \showISBNx    \undefined \def \showISBNx     #1{\unskip}     \fi
\ifx \showISBNxiii \undefined \def \showISBNxiii  #1{\unskip}     \fi
\ifx \showISSN     \undefined \def \showISSN      #1{\unskip}     \fi
\ifx \showLCCN     \undefined \def \showLCCN      #1{\unskip}     \fi
\ifx \shownote     \undefined \def \shownote      #1{#1}          \fi
\ifx \showarticletitle \undefined \def \showarticletitle #1{#1}   \fi
\ifx \showURL      \undefined \def \showURL       #1{#1}          \fi

\bibitem[\protect\citeauthoryear{Akerlof and Shiller}{Akerlof and
  Shiller}{2015}]%
        {akerlof2015phishing}
{George~A Akerlof} {and} {Robert~J Shiller}. 2015.
\newblock {\em Phishing for phools: The economics of manipulation and
  deception}.
\newblock Princeton U. Press.
\newblock


\bibitem[\protect\citeauthoryear{Alpcan and Basar}{Alpcan and Basar}{2003}]%
        {Alpcan2003}
{Tansu Alpcan} {and} {Tamer Basar}. 2003.
\newblock \showarticletitle{A game theoretic approach to decision and analysis
  in network intrusion detection}. In {\em IEEE Conf. Decision and Control},
  Vol.~3. IEEE, 2595--2600.
\newblock


\bibitem[\protect\citeauthoryear{Alvim, Chatzikokolakis, Kawamoto, and
  Palamidessi}{Alvim et~al\mbox{.}}{2017}]%
        {alvim2017leakage}
{M\'{a}rio~S. Alvim}, {Konstantinos Chatzikokolakis}, {Yusuke Kawamoto}, {and}
  {Catuscia Palamidessi}. 2017.
\newblock \showarticletitle{Information Leakage Games}. In {\em Decision and
  Game Theory for Security}. Springer, 437--457.
\newblock


\bibitem[\protect\citeauthoryear{Astyk, Newton, and Camerer}{Astyk
  et~al\mbox{.}}{2010}]%
        {astyk2010pinocchio}
{Sharon Astyk}, {Aaron Newton}, {and} {Colin~F Camerer}. 2010.
\newblock \showarticletitle{Pinocchio's pupil: using eyetracking and pupil
  dilation to understand truth telling and deception in sender-receiver games}.
\newblock {\em The American Economic Review\/} {100}, 3 (2010), 984--1007.
\newblock


\bibitem[\protect\citeauthoryear{Basar}{Basar}{1983}]%
        {basar1983gaussian}
{Tamer Basar}. 1983.
\newblock \showarticletitle{The {Gauss}ian test channel with an intelligent
  jammer}.
\newblock {\em IEEE Transactions on Information Theory\/} {29}, 1 (1983),
  152--157.
\newblock


\bibitem[\protect\citeauthoryear{Bell and Whaley}{Bell and Whaley}{2017}]%
        {bell2017cheating}
{J~Bowyer Bell} {and} {Barton Whaley}. 2017.
\newblock {\em Cheating and deception}.
\newblock Routledge.
\newblock


\bibitem[\protect\citeauthoryear{Bennett and Waltz}{Bennett and Waltz}{2007}]%
        {bennett2007counterdeception}
{Michael Bennett} {and} {Edward Waltz}. 2007.
\newblock {\em Counterdeception principles and applications for national
  security}.
\newblock Artech House.
\newblock


\bibitem[\protect\citeauthoryear{Bodmer, Kilger, Carpenter, and Jones}{Bodmer
  et~al\mbox{.}}{2012}]%
        {bodmer2012reverse}
{Sean Bodmer}, {Max Kilger}, {Gregory Carpenter}, {and} {Jade Jones}. 2012.
\newblock {\em Reverse deception: organized cyber threat counter-exploitation}.
\newblock McGraw Hill Professional.
\newblock


\bibitem[\protect\citeauthoryear{Bond~Jr and DePaulo}{Bond~Jr and
  DePaulo}{2008}]%
        {bond2008individual}
{Charles~F Bond~Jr} {and} {Bella~M DePaulo}. 2008.
\newblock \showarticletitle{Individual differences in judging deception:
  accuracy and bias.}
\newblock {\em Psychological Bulletin\/} {134}, 4 (2008), 477.
\newblock


\bibitem[\protect\citeauthoryear{Carroll and Grosu}{Carroll and Grosu}{2011}]%
        {Carroll2011}
{Thomas~E Carroll} {and} {Daniel Grosu}. 2011.
\newblock \showarticletitle{A game theoretic investigation of deception in
  network security}.
\newblock {\em Security and Commun. Nets.\/} {4}, 10 (2011), 1162--1172.
\newblock


\bibitem[\protect\citeauthoryear{{\c{C}}eker, Zhuang, Upadhyaya, La, and
  Soong}{{\c{C}}eker et~al\mbox{.}}{2016}]%
        {ceker2016deception}
{Hayreddin {\c{C}}eker}, {Jun Zhuang}, {Shambhu Upadhyaya}, {Quang~Duy La},
  {and} {Boon-Hee Soong}. 2016.
\newblock \showarticletitle{Deception-based game theoretical approach to
  mitigate DoS attacks}. In {\em Decision and Game Theory for Security}.
  Springer, 18--38.
\newblock


\bibitem[\protect\citeauthoryear{Chessa, Grossklags, and Loiseau}{Chessa
  et~al\mbox{.}}{2015}]%
        {chessa2015game}
{Michela Chessa}, {Jens Grossklags}, {and} {Patrick Loiseau}. 2015.
\newblock \showarticletitle{A game-theoretic study on non-monetary incentives
  in data analytics projects with privacy implications}. In {\em IEEE Comput.
  Security Foundations Symp.} 90--104.
\newblock


\bibitem[\protect\citeauthoryear{Chisholm}{Chisholm}{1911}]%
        {chisholm1911taxonomy}
{Hugh Chisholm}. 1911.
\newblock \showarticletitle{Predicables}.
\newblock In {\em Encyclopedia Britannica (11th edition)}. Cambridge University
  Press.
\newblock


\bibitem[\protect\citeauthoryear{Clark, Zhu, Poovendran, and Ba{\c{s}}ar}{Clark
  et~al\mbox{.}}{2012}]%
        {clark2012deceptive}
{Andrew Clark}, {Quanyan Zhu}, {Radha Poovendran}, {and} {Tamer Ba{\c{s}}ar}.
  2012.
\newblock \showarticletitle{Deceptive routing in relay networks}. In {\em
  Decision and Game Theory for Security}. Springer, 171--185.
\newblock


\bibitem[\protect\citeauthoryear{Cott}{Cott}{1940}]%
        {cott1940adaptive}
{Hugh Cott}. 1940.
\newblock {\em Adaptive Coloration in Animals}.
\newblock Methuen. 397 pages.
\newblock


\bibitem[\protect\citeauthoryear{Crawford and Sobel}{Crawford and
  Sobel}{1982}]%
        {crawford1982strategic}
{Vincent~P Crawford} {and} {Joel Sobel}. 1982.
\newblock \showarticletitle{Strategic information transmission}.
\newblock {\em Econometrica: J of the Econometric Soc.\/} (1982), 1431--1451.
\newblock


\bibitem[\protect\citeauthoryear{Do, Tran, Hong, Kamhoua, Kwiat, Blasch, Ren,
  Pissinou, and Iyengar}{Do et~al\mbox{.}}{2017}]%
        {do2017game}
{Cuong~T Do}, {Nguyen~H Tran}, {Choongseon Hong}, {Charles~A Kamhoua}, {Kevin~A
  Kwiat}, {Erik Blasch}, {Shaolei Ren}, {Niki Pissinou}, {and}
  {Sundaraja~Sitharama Iyengar}. 2017.
\newblock \showarticletitle{Game Theory for Cyber Security and Privacy}.
\newblock {\em ACM Computing Surveys (CSUR)\/} {50}, 2 (2017), 30.
\newblock


\bibitem[\protect\citeauthoryear{Durkota, Lis{\`y}, Bo\v{s}ansk{\'y}, and
  Kiekintveld}{Durkota et~al\mbox{.}}{2015}]%
        {durkota2015optimal}
{Karel Durkota}, {Viliam Lis{\`y}}, {Branislav Bo\v{s}ansk{\'y}}, {and}
  {Christopher Kiekintveld}. 2015.
\newblock \showarticletitle{Optimal Network Security Hardening Using Attack
  Graph Games}. In {\em Intl. Joint Conf. on Artificial Intelligence}.
  526--532.
\newblock


\bibitem[\protect\citeauthoryear{Edwards, Hofmeyr, and Forrest}{Edwards
  et~al\mbox{.}}{2016}]%
        {edwards2016hype}
{Benjamin Edwards}, {Steven Hofmeyr}, {and} {Stephanie Forrest}. 2016.
\newblock \showarticletitle{Hype and heavy tails: A closer look at data
  breaches}.
\newblock {\em J Cybersecurity\/} {2}, 1 (2016), 3--14.
\newblock


\bibitem[\protect\citeauthoryear{Feng, Zheng, Mohapatra, and Cansever}{Feng
  et~al\mbox{.}}{2017}]%
        {feng2017mtd}
{Xiaotao Feng}, {Zizhan Zheng}, {Prasant Mohapatra}, {and} {Derya Cansever}.
  2017.
\newblock \showarticletitle{A {Stackelberg} Game and {Markov} Modeling of
  Moving Target Defense}. In {\em Decision and Game Theory for Security}.
  Springer, 315--335.
\newblock


\bibitem[\protect\citeauthoryear{Ferguson-Walter, LaFon, and
  Shade}{Ferguson-Walter et~al\mbox{.}}{2017}]%
        {fergusonwalter2017friend}
{KJ Ferguson-Walter}, {DS LaFon}, {and} {TB Shade}. 2017.
\newblock \showarticletitle{Friend or <em>Faux</em>: Deception for Cyber
  Defense}.
\newblock {\em Journal of Information Warfare\/} {16}, 2 (2017), 28--42.
\newblock
\showURL{%
\url{https://www.jstor.org/stable/26502755}}


\bibitem[\protect\citeauthoryear{Filar and Vrieze}{Filar and Vrieze}{2012}]%
        {filar2012competitive}
{Jerzy Filar} {and} {Koos Vrieze}. 2012.
\newblock {\em Competitive Markov Decision Processes}.
\newblock Springer Science \& Business Media, New York.
\newblock


\bibitem[\protect\citeauthoryear{Fischbacher and F{\"o}llmi-Heusi}{Fischbacher
  and F{\"o}llmi-Heusi}{2013}]%
        {fischbacher2013lies}
{Urs Fischbacher} {and} {Franziska F{\"o}llmi-Heusi}. 2013.
\newblock \showarticletitle{Lies in disguise-an experimental study on
  cheating}.
\newblock {\em J European Econ. Assoc.\/} {11}, 3 (2013), 525--547.
\newblock


\bibitem[\protect\citeauthoryear{Freudiger, Manshaei, Hubaux, and
  Parkes}{Freudiger et~al\mbox{.}}{2009}]%
        {freudiger2009loc}
{Julien Freudiger}, {Mohammad~Hossein Manshaei}, {Jean-Pierre Hubaux}, {and}
  {David~C Parkes}. 2009.
\newblock \showarticletitle{On non-cooperative location privacy: a
  game-theoretic analysis}. In {\em Proc. ACM Conf. on Comput. and Commun.
  Security}. ACM, 324--337.
\newblock


\bibitem[\protect\citeauthoryear{Fudenberg and Tirole}{Fudenberg and
  Tirole}{1991}]%
        {fudenberg1991game}
{D. Fudenberg} {and} {J. Tirole}. 1991.
\newblock {\em Game Theory}. Vol. 393.
\newblock


\bibitem[\protect\citeauthoryear{Geiselman}{Geiselman}{2012}]%
        {geiselman2012cognitive}
{R~Edward Geiselman}. 2012.
\newblock \showarticletitle{The cognitive interview for suspects (CIS)}.
\newblock {\em American College of Forensic Psychology\/} {30}, 3 (2012),
  1--16.
\newblock


\bibitem[\protect\citeauthoryear{Gneezy}{Gneezy}{2005}]%
        {gneezy2005deception}
{Uri Gneezy}. 2005.
\newblock \showarticletitle{Deception: The role of consequences}.
\newblock {\em American Econ. Review\/} {95}, 1 (2005), 384--394.
\newblock


\bibitem[\protect\citeauthoryear{Godson and Wirtz}{Godson and Wirtz}{2011}]%
        {godson2011strategic}
{Roy Godson} {and} {James~J Wirtz}. 2011.
\newblock {\em Strategic denial and deception: the Twenty-first Century
  challenge}.
\newblock Transaction Publishers.
\newblock


\bibitem[\protect\citeauthoryear{Grosser}{Grosser}{2014}]%
        {grosser2014scare}
{Benjamin Grosser}. 2014.
\newblock \showarticletitle{Privacy through visibility: disrupting {NSA}
  Surveillance with Algorithmically Generated "Scary" Stories}. University of
  Wisconsin-Milwaukee.
\newblock
\showURL{%
\url{https://bengrosser.com/projects/scaremail/}}


\bibitem[\protect\citeauthoryear{Heckman, Stech, Thomas, Schmoker, and
  Tsow}{Heckman et~al\mbox{.}}{2015}]%
        {heckman2015cyber}
{Kristin~E Heckman}, {Frank~J Stech}, {Roshan~K Thomas}, {Ben Schmoker}, {and}
  {Alexander~W Tsow}. 2015.
\newblock {\em Cyber denial, deception and counter deception}.
\newblock Springer.
\newblock


\bibitem[\protect\citeauthoryear{Hor\'{a}k, Zhu, and
  Bo\v{s}ansk{\'y}}{Hor\'{a}k et~al\mbox{.}}{2017}]%
        {horak2017manipulating}
{Karel Hor\'{a}k}, {Quanyan Zhu}, {and} {Branislav Bo\v{s}ansk{\'y}}. 2017.
\newblock \showarticletitle{Manipulating Adversary's Belief: A Dynamic Game
  Approach to Deception by Design in Network Security}. In {\em Decision and
  Game Theory for Security}. Springer, 273--294.
\newblock


\bibitem[\protect\citeauthoryear{Howe and Nissenbaum}{Howe and
  Nissenbaum}{2009}]%
        {Howe2009}
{Daniel~C. Howe} {and} {Helen Nissenbaum}. 2009.
\newblock \showarticletitle{{TrackMeNot}: Resisting surveillance in web
  search}.
\newblock {\em Lessons from the Identity Trail: Anonymity, Privacy, and
  Identity in a Networked Society\/}  {23} (2009), 417--436.
\newblock
\showURL{%
\url{http://www.nyu.edu/pages/projects/nissenbaum/papers/ch23(HoweNissenbaum)Web.pdf}}


\bibitem[\protect\citeauthoryear{Hurkens and Kartik}{Hurkens and
  Kartik}{2009}]%
        {hurkens2009would}
{Sjaak Hurkens} {and} {Navin Kartik}. 2009.
\newblock \showarticletitle{Would {I} lie to you? On social preferences and
  lying aversion}.
\newblock {\em Experimental Economics\/} {12}, 2 (2009), 180--192.
\newblock


\bibitem[\protect\citeauthoryear{Jain, Tsai, Pita, Kiekintveld, Rathi, Tambe,
  and Ord{\'o}nez}{Jain et~al\mbox{.}}{2010}]%
        {jain2010software}
{Manish Jain}, {Jason Tsai}, {James Pita}, {Christopher Kiekintveld},
  {Shyamsunder Rathi}, {Milind Tambe}, {and} {Fernando Ord{\'o}nez}. 2010.
\newblock \showarticletitle{Software assistants for randomized patrol planning
  for the {LAX} airport police and the {Federal} {Air} {Marshal} {Service}}.
\newblock {\em Interfaces\/} {40}, 4 (2010), 267--290.
\newblock


\bibitem[\protect\citeauthoryear{Kartik}{Kartik}{2009}]%
        {kartik2009strategic}
{Navin Kartik}. 2009.
\newblock \showarticletitle{Strategic communication with lying costs}.
\newblock {\em The Review of Economic Studies\/} {76}, 4 (2009), 1359--1395.
\newblock


\bibitem[\protect\citeauthoryear{Kiekintveld, Lis{\`y}, and
  P{\'\i}bil}{Kiekintveld et~al\mbox{.}}{2015}]%
        {kiekintveld2015game}
{Christopher Kiekintveld}, {Viliam Lis{\`y}}, {and} {Radek P{\'\i}bil}. 2015.
\newblock \showarticletitle{Game-theoretic foundations for the strategic use of
  honeypots in network security}.
\newblock In {\em Cyber Warfare}. Springer, 81--101.
\newblock


\bibitem[\protect\citeauthoryear{Lu, Lin, Luan, Liang, and Shen}{Lu
  et~al\mbox{.}}{2012}]%
        {lu2012pseudonym}
{Rongxing Lu}, {Xiaodong Lin}, {Tom~H Luan}, {Xiaohui Liang}, {and} {Xuemin
  Shen}. 2012.
\newblock \showarticletitle{Pseudonym changing at social spots: An effective
  strategy for location privacy in vanets}.
\newblock {\em IEEE Trans Vehicular Technol.\/} {61}, 1 (2012), 86--96.
\newblock


\bibitem[\protect\citeauthoryear{Lykken}{Lykken}{1959}]%
        {lykken1959gsr}
{David~T Lykken}. 1959.
\newblock \showarticletitle{The {GSR} in the detection of guilt}.
\newblock {\em Journal of Applied Psychology\/} {43}, 6 (1959), 385.
\newblock


\bibitem[\protect\citeauthoryear{Mahon}{Mahon}{2016}]%
        {stanford2016deception}
{James~Edwin Mahon}. 2016.
\newblock \showarticletitle{The Definition of Lying and Deception}.
\newblock In {\em The Stanford Encyclopedia of Philosophy} (winter 2016 ed.),
  {Edward~N. Zalta} (Ed.).
\newblock


\bibitem[\protect\citeauthoryear{Manshaei, Zhu, Alpcan, Bac{\c{s}}ar, and
  Hubaux}{Manshaei et~al\mbox{.}}{2013}]%
        {manshaei2013game}
{Mohammad~Hossein Manshaei}, {Quanyan Zhu}, {Tansu Alpcan}, {Tamer
  Bac{\c{s}}ar}, {and} {Jean-Pierre Hubaux}. 2013.
\newblock \showarticletitle{Game theory meets network security and privacy}.
\newblock {\em ACM Computing Surveys (CSUR)\/} {45}, 3 (2013), 25.
\newblock


\bibitem[\protect\citeauthoryear{Meyerowitz and Roy~Choudhury}{Meyerowitz and
  Roy~Choudhury}{2009}]%
        {meyerowitzhiding2009}
{Joseph Meyerowitz} {and} {Romit Roy~Choudhury}. 2009.
\newblock \showarticletitle{Hiding stars with fireworks: location privacy
  through camouflage}. In {\em Proceedings of the 15\textsuperscript{th} annual
  Intl. Conf. on {Mobile} computing and networking}. ACM, 345--356.
\newblock
\showURL{%
\url{http://dl.acm.org/citation.cfm?id=1614358}}


\bibitem[\protect\citeauthoryear{Milgrom}{Milgrom}{1981}]%
        {milgrom1981good}
{Paul~R Milgrom}. 1981.
\newblock \showarticletitle{Good news and bad news: Representation theorems and
  applications}.
\newblock {\em Bell J. of Econ.\/} (1981), 380--391.
\newblock


\bibitem[\protect\citeauthoryear{MITRE}{MITRE}{2010}]%
        {mitre2010science}
{MITRE}. 2010.
\newblock \showarticletitle{Science of Cyber-Security}.
\newblock  (2010).
\newblock


\bibitem[\protect\citeauthoryear{Mohammadi, Manshaei, Moghaddam, and
  Zhu}{Mohammadi et~al\mbox{.}}{2016}]%
        {mohammadi2016game}
{Amin Mohammadi}, {Mohammad~Hossein Manshaei}, {Monireh~Mohebbi Moghaddam},
  {and} {Quanyan Zhu}. 2016.
\newblock \showarticletitle{A Game-Theoretic Analysis of Deception over Social
  Networks Using Fake Avatars}. In {\em Decision and Game Theory for Security}.
  Springer, 382--394.
\newblock


\bibitem[\protect\citeauthoryear{Myerson}{Myerson}{1991}]%
        {myerson1991gametheory}
{Roger~B. Myerson}. 1991.
\newblock {\em Game Theory: Analysis of Conflict}.
\newblock Harvard U. Press.
\newblock


\bibitem[\protect\citeauthoryear{Nash}{Nash}{1950}]%
        {nash1950equilibrium}
{John~F Nash}. 1950.
\newblock \showarticletitle{Equilibrium points in n-person games}.
\newblock {\em Proc. Nat. Acad. Sci. USA\/} {36}, 1 (1950), 48--49.
\newblock


\bibitem[\protect\citeauthoryear{NISO}{NISO}{2005}]%
        {niso2005vocabularies}
{NISO}. 2005.
\newblock \showarticletitle{Guidelines for the Construction, Format, and
  Management of Monolingual Controlled Vocabularies}.
\newblock  (2005).
\newblock


\bibitem[\protect\citeauthoryear{Nissenbaum}{Nissenbaum}{2004}]%
        {nissenbaum2004privacy}
{Helen Nissenbaum}. 2004.
\newblock \showarticletitle{Privacy as contextual integrity}.
\newblock {\em Wash. L. Rev.\/}  {79} (2004), 119.
\newblock


\bibitem[\protect\citeauthoryear{Oltramari, Cranor, Walls, and
  McDaniel}{Oltramari et~al\mbox{.}}{2014}]%
        {oltramari2014building}
{Alessandro Oltramari}, {Lorrie~Faith Cranor}, {Robert~J Walls}, {and}
  {Patrick~D McDaniel}. 2014.
\newblock \showarticletitle{Building an Ontology of Cyber Security.}. In {\em
  Conf. on Semantic Technol. for Defense, Intelligence, and Security (STIDS)}.
  54--61.
\newblock


\bibitem[\protect\citeauthoryear{Pawlick, Colbert, and Zhu}{Pawlick
  et~al\mbox{.}}{2018}]%
        {pawlick2018modeling}
{Jeffrey Pawlick}, {Edward Colbert}, {and} {Quanyan Zhu}. 2018.
\newblock \showarticletitle{Modeling and analysis of leaky deception using
  signaling games with evidence}.
\newblock {\em IEEE Trans. Inform. Forensics and Security\/} (2018).
\newblock


\bibitem[\protect\citeauthoryear{Pawlick, Farhang, and Zhu}{Pawlick
  et~al\mbox{.}}{2015}]%
        {pawlick2015flip}
{Jeffrey Pawlick}, {Sadegh Farhang}, {and} {Quanyan Zhu}. 2015.
\newblock \showarticletitle{Flip the Cloud: Cyber-Physical Signaling Games in
  the Presence of Advanced Persistent Threats}.
\newblock In {\em Decision and Game Theory for Security}. Springer, 289--308.
\newblock


\bibitem[\protect\citeauthoryear{Pawlick and Zhu}{Pawlick and Zhu}{2015}]%
        {pawlick2015deception}
{Jeffrey Pawlick} {and} {Quanyan Zhu}. 2015.
\newblock \showarticletitle{Deception by design: evidence-based signaling games
  for network defense}. In {\em Workshop on the Economics of Inform. Security
  and Privacy}. Delft, The Netherlands.
\newblock
\showURL{%
\url{http://arxiv.org/abs/1503.05458}}


\bibitem[\protect\citeauthoryear{Pawlick and Zhu}{Pawlick and Zhu}{2016}]%
        {pawlick2016stackelberg}
{Jeffrey Pawlick} {and} {Quanyan Zhu}. 2016.
\newblock \showarticletitle{A {Stackelberg} Game Perspective on the Conflict
  Between Machine Learning and Data Obfuscation}. In {\em IEEE Workshop on
  Inform. Forensics and Security}.
\newblock
\showURL{%
\url{https://arxiv.org/abs/1608.02546}}


\bibitem[\protect\citeauthoryear{Pawlick and Zhu}{Pawlick and Zhu}{2017a}]%
        {pawlick2017meanfield}
{Jeffrey Pawlick} {and} {Quanyan Zhu}. 2017a.
\newblock \showarticletitle{A Mean-Field {Stackelberg} Game Approach for
  Obfuscation Adoption in Empirical Risk Minimization}. In {\em Global Signal
  and Inform. Processing Workshop on Control and Game Theoretic Approaches to
  Security and Privacy}.
\newblock


\bibitem[\protect\citeauthoryear{Pawlick and Zhu}{Pawlick and Zhu}{2017b}]%
        {pawlick2017strategic}
{Jeffrey Pawlick} {and} {Quanyan Zhu}. 2017b.
\newblock \showarticletitle{Strategic Trust in Cloud-Enabled Cyber-Physical
  Systems with an Application to Glucose Control}.
\newblock {\em IEEE Trans Inform. Forensics and Security\/} {12}, 12 (2017),
  2906--2919.
\newblock


\bibitem[\protect\citeauthoryear{Peppet}{Peppet}{2014}]%
        {peppetregulating2014}
{Scott~R. Peppet}. 2014.
\newblock \showarticletitle{Regulating the {Internet} of Things: First Steps
  toward Managing Discrimination, Privacy, Security and Consent}.
\newblock {\em Tex. L. Rev.\/}  {93} (2014), 85.
\newblock
\showURL{%
\url{http://heinonlinebackup.com/hol-cgi-bin/getpdf.cgi?handle=hein.J.s/tlr93&section=5}}


\bibitem[\protect\citeauthoryear{P\'{i}bil, Lis{\`y}, Kiekintveld,
  Bo{\v{s}}ansk{\`y}, and Pechoucek}{P\'{i}bil et~al\mbox{.}}{2012}]%
        {pibil2012game}
{Radek P\'{i}bil}, {Viliam Lis{\`y}}, {Christopher Kiekintveld}, {Branislav
  Bo{\v{s}}ansk{\`y}}, {and} {Michal Pechoucek}. 2012.
\newblock \showarticletitle{Game theoretic model of strategic honeypot
  selection in computer networks}. In {\em Decision and Game Theory for
  Security}. Springer, 201--220.
\newblock


\bibitem[\protect\citeauthoryear{Pita, Jain, Marecki, Ord{\'o}{\~n}ez, Portway,
  Tambe, Western, Paruchuri, and Kraus}{Pita et~al\mbox{.}}{2008}]%
        {pita2008deployed}
{James Pita}, {Manish Jain}, {Janusz Marecki}, {Fernando Ord{\'o}{\~n}ez},
  {Christopher Portway}, {Milind Tambe}, {Craig Western}, {Praveen Paruchuri},
  {and} {Sarit Kraus}. 2008.
\newblock \showarticletitle{Deployed ARMOR protection: the application of a
  game theoretic model for security at the Los Angeles Intl. Airport}. In {\em
  Proceedings of the 7\textsuperscript{th} Intl. joint Conf. on Autonomous
  agents and multiagent systems: industrial track}. Intl. Foundation for
  Autonomous Agents and Multiagent Systems, 125--132.
\newblock


\bibitem[\protect\citeauthoryear{Rass, K\"{o}nig, and Schauer}{Rass
  et~al\mbox{.}}{2017}]%
        {rass2017controlmixed}
{Stefan Rass}, {Sandra K\"{o}nig}, {and} {Stefan Schauer}. 2017.
\newblock \showarticletitle{On the Cost of Game Playing: How to Control the
  Expenses in Mixed Strategies}. In {\em Decision and Game Theory for
  Security}. Springer, 495--505.
\newblock


\bibitem[\protect\citeauthoryear{Rothstein and Whaley}{Rothstein and
  Whaley}{2013}]%
        {rothstein2013art}
{Hy Rothstein} {and} {Barton Whaley}. 2013.
\newblock {\em The art and science of military deception}.
\newblock Artech House.
\newblock


\bibitem[\protect\citeauthoryear{Rowe}{Rowe}{2006}]%
        {rowe2006taxonomy}
{Neil~C Rowe}. 2006.
\newblock \showarticletitle{A taxonomy of deception in cyberspace}.
\newblock {\em Intl. Conf. Inform. Warfare and Security\/} (2006).
\newblock


\bibitem[\protect\citeauthoryear{Rowe and Rrushi}{Rowe and Rrushi}{2016}]%
        {rowe2016introduction}
{Neil~C Rowe} {and} {Julian Rrushi}. 2016.
\newblock {\em Introduction to Cyberdeception}.
\newblock Springer.
\newblock


\bibitem[\protect\citeauthoryear{Roy, Ellis, Shiva, Dasgupta, Shandilya, and
  Wu}{Roy et~al\mbox{.}}{2010}]%
        {roy2010survey}
{Sankardas Roy}, {Charles Ellis}, {Sajjan Shiva}, {Dipankar Dasgupta}, {Vivek
  Shandilya}, {and} {Qishi Wu}. 2010.
\newblock \showarticletitle{A survey of game theory as applied to network
  security}. In {\em IEEE Intl. Conf. on System Sciences}. 1--10.
\newblock


\bibitem[\protect\citeauthoryear{Sengupta, Chowdhary, Huang, and
  Kambhampati}{Sengupta et~al\mbox{.}}{2018}]%
        {sengupta2018moving}
{Sailik Sengupta}, {Ankur Chowdhary}, {Dijiang Huang}, {and} {Subbarao
  Kambhampati}. 2018.
\newblock \showarticletitle{Moving target defense for the placement of
  intrusion detection systems in the cloud}. In {\em Decision and Game Theory
  for Security}. Springer, 326--345.
\newblock


\bibitem[\protect\citeauthoryear{Shieh, An, Yang, Tambe, Baldwin, DiRenzo,
  Maule, and Meyer}{Shieh et~al\mbox{.}}{2012}]%
        {shieh2012protect}
{Eric Shieh}, {Bo An}, {Rong Yang}, {Milind Tambe}, {Craig Baldwin}, {Joseph
  DiRenzo}, {Ben Maule}, {and} {Garrett Meyer}. 2012.
\newblock \showarticletitle{Protect: A deployed game theoretic system to
  protect the ports of the {United}{States}}. In {\em Proceedings of the
  11\textsuperscript{th} Intl. Conf. on Autonomous Agents and Multiagent
  Systems-Volume 1}. Intl. Foundation for Autonomous Agents and Multiagent
  Systems, 13--20.
\newblock


\bibitem[\protect\citeauthoryear{Shokri}{Shokri}{2015}]%
        {shokri2015privacy}
{Reza Shokri}. 2015.
\newblock \showarticletitle{Privacy games: Optimal user-centric data
  obfuscation}.
\newblock {\em Proc. Privacy Enhancing Technologies\/} 2 (2015), 299--315.
\newblock


\bibitem[\protect\citeauthoryear{Theodorakopoulos, Shokri, Troncoso, Hubaux,
  and Le~Boudec}{Theodorakopoulos et~al\mbox{.}}{2014}]%
        {theodorakopoulos2014prolonging}
{George Theodorakopoulos}, {Reza Shokri}, {Carmela Troncoso}, {Jean-Pierre
  Hubaux}, {and} {Jean-Yves Le~Boudec}. 2014.
\newblock \showarticletitle{Prolonging the hide-and-seek game: Optimal
  trajectory privacy for location-based services}. In {\em Proc. ACM Workshop
  on Privacy in the Electronic Society}. 73--82.
\newblock


\bibitem[\protect\citeauthoryear{Von~Stackelberg}{Von~Stackelberg}{1934}]%
        {vonstackelbergmarktform1934}
{Heinrich Von~Stackelberg}. 1934.
\newblock {\em Marktform und gleichgewicht}.
\newblock J. Springer.
\newblock


\bibitem[\protect\citeauthoryear{Vrij, Mann, Fisher, Leal, Milne, and
  Bull}{Vrij et~al\mbox{.}}{2008}]%
        {vrij2008increasing}
{Aldert Vrij}, {Samantha~A Mann}, {Ronald~P Fisher}, {Sharon Leal}, {Rebecca
  Milne}, {and} {Ray Bull}. 2008.
\newblock \showarticletitle{Increasing cognitive load to facilitate lie
  detection: The benefit of recalling an event in reverse order}.
\newblock {\em Law and Human Behavior\/} {32}, 3 (2008), 253--265.
\newblock


\bibitem[\protect\citeauthoryear{Whaley}{Whaley}{2016}]%
        {whaley2016practise}
{Barton Whaley}. 2016.
\newblock {\em Practise to Deceive: Learning Curves of Military Deception
  Planners}.
\newblock Naval Institute Press.
\newblock


\bibitem[\protect\citeauthoryear{Zhang, Yu, Fu, and Das}{Zhang
  et~al\mbox{.}}{2010}]%
        {zhang2010gpath}
{Nan Zhang}, {Wei Yu}, {Xinwen Fu}, {and} {Sajal~K Das}. 2010.
\newblock \showarticletitle{g{P}ath: A Game-Theoretic Path Selection Algorithm
  to Protect Tor's Anonymity}. In {\em Decision and Game Theory for Security}.
  Springer, 58--71.
\newblock


\bibitem[\protect\citeauthoryear{Zhang and Zhu}{Zhang and Zhu}{2015}]%
        {zhang2015secure}
{Rui Zhang} {and} {Quanyan Zhu}. 2015.
\newblock \showarticletitle{Secure and resilient distributed machine learning
  under adversarial environments}. In {\em Information Fusion (Fusion), 2015
  18th International Conference on}. IEEE, 644--651.
\newblock


\bibitem[\protect\citeauthoryear{Zhang and Zhu}{Zhang and Zhu}{2017}]%
        {zhang2017game}
{Rui Zhang} {and} {Quanyan Zhu}. 2017.
\newblock \showarticletitle{A game-theoretic analysis of label flipping attacks
  on distributed support vector machines}. In {\em Information Sciences and
  Systems (CISS), 2017 51st Annual Conference on}. IEEE, 1--6.
\newblock


\bibitem[\protect\citeauthoryear{Zhu and Ba{\c{s}}ar}{Zhu and
  Ba{\c{s}}ar}{2013}]%
        {zhu2013game}
{Quanyan Zhu} {and} {Tamer Ba{\c{s}}ar}. 2013.
\newblock \showarticletitle{Game-theoretic approach to feedback-driven
  multi-stage moving target defense}. In {\em Decision and Game Theory for
  Security}. Springer, 246--263.
\newblock


\bibitem[\protect\citeauthoryear{Zhu, Clark, Poovendran, and Ba{\c{s}}ar}{Zhu
  et~al\mbox{.}}{2012}]%
        {zhu2012deceptive}
{Quanyan Zhu}, {Andrew Clark}, {Radha Poovendran}, {and} {Tamer Ba{\c{s}}ar}.
  2012.
\newblock \showarticletitle{Deceptive routing games}. In {\em IEEE Conf. on
  Decision and Control}. 2704--2711.
\newblock


\bibitem[\protect\citeauthoryear{Zhuang, Bier, and Alagoz}{Zhuang
  et~al\mbox{.}}{2010}]%
        {zhuang2010modeling}
{J. Zhuang}, {V.~M. Bier}, {and} {O. Alagoz}. 2010.
\newblock \showarticletitle{Modeling secrecy and deception in a multiple-period
  attacker--defender signaling game}.
\newblock {\em European J Operational Res.\/} {203}, 2 (2010), 409--418.
\newblock


\end{thebibliography}

\end{document}